\documentclass{SciPost}

\usepackage[bitstream-charter]{mathdesign}
\usepackage{mathtools}
\usepackage{braket}
\usepackage{slashed}
\usepackage{xcolor}
\usepackage{comment}

\hypersetup{
    colorlinks,
    linkcolor={red!50!black},
    citecolor={blue!50!black},
    urlcolor={blue!80!black}
}

\DeclareSymbolFont{usualmathcal}{OMS}{cmsy}{m}{n}
\DeclareSymbolFontAlphabet{\mathcal}{usualmathcal}

\fancypagestyle{SPstyle}{
\fancyhf{}
\lhead{\colorbox{scipostblue}{\bf \color{white} ~SciPost Physics }}
\rhead{{\bf \color{scipostdeepblue} ~Submission }}

\fancyfoot[C]{\textbf{\thepage}}
}

\usepackage{maksmath}

\usepackage{marginnote}
\usepackage{ulem}

\renewcommand{\emph}[1]{\textit{#1}}

\newcommand{\Sgrad}{S^{\text{grad}}}
\newcommand{\Stop}{S^{\text{top}}}

\newcommand{\ltop}{\epsilon_{\mu\nu} \tr \, (Q\del_\mu Q \del_\nu Q)}

\begin{document}

\pagestyle{SPstyle}

\begin{center}{\Large \textbf{\color{scipostdeepblue}{
Continuum field theory of matchgate tensor network ensembles
}}}\end{center}

\begin{center}\textbf{
Maksimilian Usoltcev\textsuperscript{1$\star$},
Carolin Wille\textsuperscript{2$\ast$},
Jens Eisert\textsuperscript{3,4$\dagger$} and
Alexander Altland\textsuperscript{1$\ddagger$}
}\end{center}

\begin{center}
{\small
{\bf 1} Institut f\"ur Theoretische Physik, 50937 Cologne, Germany
\\
{\bf 2} Department of Applied Mathematics and Theoretical Physics, University of Cambridge,\\
Cambridge CB3 0WA, United Kingdom 
\\
{\bf 3} Dahlem Center for Complex Quantum Systems, Freie Universit{\"a}t
Berlin, 14195 Berlin, Germany
\\
{\bf 4}
Helmholtz-Zentrum Berlin f\"ur Materialien und Energie, 14109 Berlin, Germany
}
\\[\baselineskip]
$\star$ \href{mailto:usoltcev@thp.uni-koeln.de}{\small usoltcev@thp.uni-koeln.de} \\
$\ast$ \href{mailto:carolin.wille@fu-berlin.de}{\small carolin.wille@fu-berlin.de}\\
$\dagger$ \href{mailto:jense@zedat.fu-berlin.de}{\small jense@zedat.fu-berlin.de} \\
$\ddagger$ \href{mailto:alexal@thp.uni-koeln.de}{\small alexal@thp.uni-koeln.de}
\end{center}

\section*{\color{scipostdeepblue}{Abstract}}

{\bf Tensor networks provide discrete representations of quantum many-body systems, yet their precise connection to continuum field theories remains relatively poorly understood. 
Invoking a notion of typicality, we develop a continuum
description for random \emph{ensembles} of two-dimensional fermionic matchgate tensor networks with spatially fluctuating parameters. {As a diagnostic of the resulting universal physics, we analyze disorder-averaged moments of fermionic two-point functions, both in flat geometry and on a hyperbolic disk, where curvature reshapes their long-distance structure.}
We show that disorder drives universal long-distance behavior governed by a nonlinear sigma model of symmetry class~D with a topological term,
placing random matchgate networks in direct correspondence with the thermal quantum Hall problem. 
The resulting phase structure includes localized phases, quantum Hall criticality, and a robust thermal metal with diffusive correlations and spontaneous replica-symmetry breaking. 
Weak non-Gaussian deformations reduce the symmetry to discrete
permutations, generate a mass for the Goldstone modes, and suppress long-range correlations. 
In this way, typicality offers a route from ensembles of discrete
tensor networks to continuum quantum field theories. 
}
\vspace{\baselineskip}



\vspace{10pt}
\noindent\rule{\textwidth}{1pt}
\tableofcontents
\noindent\rule{\textwidth}{1pt}
\vspace{10pt}

\section{Introduction}
\label{sec:intro}
Among the prominent approaches to capture complex many-body systems,
tensor-network-based methods and continuum field theories each have their distinct strengths: tensor-network methods are powerful numerical and analytical
tools for understanding strongly correlated discrete regimes \cite{Orus-AnnPhys-2014,RevModPhys.93.045003,Bridgeman2017,AreaReview}, while continuum
field theories capture universal behavior on large length scales \cite{Altland2023, Efetbook}. Despite their
respective successes, establishing a controlled connection between these viewpoints and approaches --- particularly in more than one spatial dimensions and away from
exactly solvable limits --- remains largely an 
open problem, despite 
recent steps in this direction \cite{PhysRevLett.104.190405,PhysRevLett.105.260401,PhysRevB.88.085118, PhysRevX.9.021040}.

In this work, we address this difficulty by adopting an ensemble-based
perspective. Rather than focusing on individual tensor-network realizations,
whose microscopic details are inherently discrete and model-dependent, we
formulate statements at the level of \emph{ensembles} and their typical 
properties.
This viewpoint enables a controlled coarse-graining in which microscopic
variability is integrated out, while structural features relevant at large
length scales --- such as symmetries, topology, and universal correlations --- are
retained. Ensemble averaging thus provides a natural interface between
tensor-network descriptions and continuum field-theoretic methods. 
In this way, methods of
\emph{random tensor networks} feature strongly \cite{PhysRevA.81.032336, Hayden2016,PRXQuantum.2.040308, LancienRandomMPS}, but here with a new twist.
{Concretely, the approach taken allows us to access disorder-averaged 
observables and their fluctuations by mapping them to correlation functions of the emergent field theory, including their dependence on 
the underlying geometry.}
{Hence, we advocate a shift in mindset away from individual realizations to a picture of \emph{typicality}, which allows us to extract continuum models from tensor networks.}

The central objects of our construction are ensembles of two-dimensional
matchgate tensor networks \cite{Cai2007OnTT,Bravyi2009,Jahn}. Matchgate tensor networks are in a one-to-one
correspondence with Gaussian (free) fermionic systems in the following sense: their contraction is given by a Gaussian fermionic Grassmann integral and hence can be computed exactly and efficiently\footnote{Efficiently here is a technical term meaning that the time needed to perform the computation grows polynomially in the number of tensors, whereas a generic exact two-dimensional tensor network contraction has exponential runtime.} via the evaluation of a Pfaffian \cite{Bravyi2009,Jahn}. 
In practice, however, the large-scale properties of matchgate tensor networks remain beyond reach within exact numerical schemes. This applies in particular to tensor networks on more complex geometries such as hyperbolic disks, where the number of tensors grows exponentially with the diameter. Even more detrimental is the inclusion of disorder. If one is interested in statements of ensembles, the sampling effort becomes prohibitive. 

It is in this regime that we propose to exploit the connection to free fermionic systems along an alternative route --- namely, as the starting point for establishing an (effective) continuum \emph{field-theoretic} description which captures the ensemble directly. Working in both flat and negatively curved geometries, we then use this formulation to derive continuum predictions for the statistics of correlation functions, including in a hyperbolic setting. 
In flat two-dimensional space, these predictions reproduce the characteristic long-distance behavior 
driven by the soft modes of disordered free fermionic systems in symmetry class D, known from the theory of disordered superconductors. On the hyperbolic geometry, curvature reshapes the long-range structure and yields boundary-dominated correlation profiles.

The effective field theory description remains applicable in the
presence of weak \emph{non-matchgate} contributions corresponding to quartic terms added to
the free fermionic theory. While such terms obstruct exact tensor-network
contractions \cite{Cai2007OnTT,Bravyi2009,Jahn}, they can be incorporated in a controlled manner at the level of
the effective field theory, where they give rise to qualitative changes with a
clear physical interpretation.

In view of the length of this work and the fact that we are addressing more than one research community --- this work attempts to introduce  the field theoretical framework in a pedagogical language accessible to non-specialists --- we start with an extended summary of results in Section~\ref{sec:SummaryOfResults}. 
{In Section~\ref{sec:GFTN}, we summarize the essential features of Gaussian fermionic tensor networks and outline their connection to superconductors in symmetry class~D. 
In Section \ref{sec:ft-derivation} we proceed to introduce the ensemble-averaged viewpoint, discuss the symmetries of the averaged model and derive its continuum description in terms of an effective field theory for a collective field $Q$ (a nonlinear $\sigma$-model). {To demonstrate its applicability, we then derive field-theoretic expressions for moments of two-point functions.}
In Section~\ref{sec:quartic}, we supplement the previously strictly Gaussian theory by controlled (weak) deviations from the Gaussian limit (non-matchgate contributions), and discuss their impact on the level of the effective long-range theory. 
{In Section~\ref{sec:hyperbolic}, we discuss the behavior of sigma-model correlators on a hyperbolic disk, elucidating how changing the geometry can modify the long-range properties of the theory.}
Finally, in Section~\ref{sec:conclusion}, we provide a brief conclusion and an outlook towards further research directions.}


\section{Summary of results}
\label{sec:SummaryOfResults}

Two-dimensional matchgate tensor networks are closely related to free fermionic
systems. Concretely, the network contraction reduces to a Gaussian Grassmann
integral 
with action $\sim \theta^\T H \theta$,
as we are 
here dealing with Majorana fermions,
where $H$ is a purely imaginary
antisymmetric matrix encoding both the tensor entries and the network
connectivity as local couplings on a lattice. Already in the simple
translation-invariant setting studied previously in
Ref.~\cite{wille2023topodual} --- a square-lattice network in which individual tensors
are determined by a single real parameter $a$ (cf.\  Fig.~\ref{fig:TNLattice}) --- this
structure gives rise to a non-trivial phase diagram organized into several
distinct topological phases.

Within each phase, the band structure of $H$, specified by eigenfunctions
$\psi_n(k)$ and eigenvalues $\eps_n(k)$, carries an integer Chern invariant
$C$ \cite{TKNN1982,QiZhang2011review}. For the minimal model considered here, four phase boundaries occur at
$a=\pm a_\pm$, separating phases with Chern numbers $C=0,-1,0,1,0$ as $a$ is tuned
from negative to positive values. Close to these transition points the system is
well described by a Haldane-Chern insulator \cite{Haldane1988,QiZhang2011review}, which in turn reduces to Dirac
fermions in the immediate vicinity of the gap-closing points.

From the perspective of free-fermion classification, purely imaginary
antisymmetric Hamiltonians belong to symmetry class~D \cite{AltlandZirnbauer,Kitaev2009PeriodicTable}. In two spatial
dimensions, this class admits topologically non-trivial insulating phases,
commonly realized in spin-triplet superconductors \cite{ReadGreen2000PairedStates}. (See Appendix
\ref{sec:phases} for a quick review of topological class~D superconductivity.) The matchgate tensor-network
Hamiltonian thus provides a concrete lattice realization of both trivial and
topological class~D phases, with the parameter $a$ controlling the topological
character of the clean system.

\subsection{Continuum field theory and disorder}

As stated before, we resort to a notion of typicality here. In this work, we go beyond individual tensor networks and study ensembles
generated by adding local Gaussian random tensors to the clean network structure
shown in Fig.~\ref{fig:TNLattice}. The width of the random-tensor distribution
defines a disorder strength $W$ and allows us to investigate how randomness
competes with the topology of the underlying clean phase. In condensed-matter
language, this setting defines the physics of the class~D thermal quantum Hall
effect~\cite{SenthilFisher2000QuasiparticleLocalization,MirlinReview,Wang2021}.

\begin{figure}
	\centering		
	\includegraphics[width=0.8\textwidth]{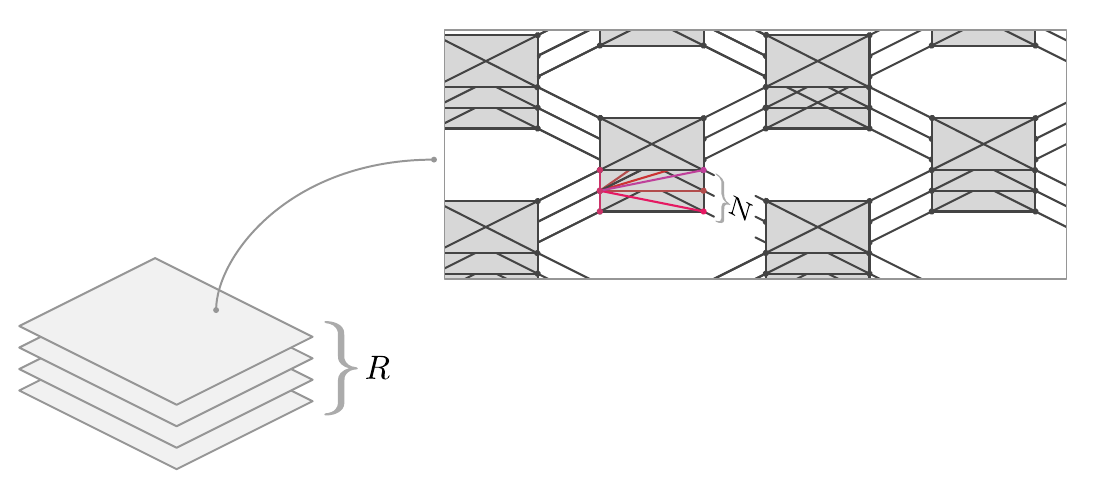}
	\caption{Setup. We consider $c=1,\ldots,N$ \emph{layers} of the clean system ($N=3$ shown above) and couple the layers with local random tensors ${\tilde A}_x$ (red). The random tensors couple all modes at a given site to all other modes at that site and do not discriminate between layers or unit-cell positions. The $N$-layer system is then replicated by $r=1,\ldots,R$ uncoupled \emph{replicas}. }
	\label{fig:layer_replica}
\end{figure}

To render the ensemble problem analytically tractable, we consider a layered
construction consisting of $N$ identical copies of the two-dimensional system,
coupled by local random tensors (cf.\ Fig.~\ref{fig:layer_replica}). From the
tensor-network perspective this corresponds to a large bond dimension, while
from a field-theoretical viewpoint it realizes a parallel shunt of weakly
conducting layers. The parameter $N$ plays the role of a control parameter in
the ensemble-averaged theory.

To describe ensemble-averaged properties, we replicate the system into $R$
identical copies (the replica trick, cf.\ Ref.~\cite{Edwards1975}). Within this approach, the Majorana
bilinear generalizes to $\sum_{r=1}^R \theta^{r\T} H \theta^r$, and a continuous
replica rotation symmetry $\theta^r \mapsto \sum_{s=1}^R T^{rs} \theta^{s}$ emerges. The fate
of this symmetry upon disorder averaging decides the universal large-scale
properties of the system.

For generic values of the control parameter $a\neq\pm a_\pm$, all energy bands
$\eps_n(k)$ remain detached from zero energy. Very weak disorder does not
qualitatively modify this situation: the ensemble-averaged spectral density at
zero energy remains vanishing, and the system stays in a band-insulating phase.
A stationary-phase analysis, formally stabilized by the large number of layers,
shows that replica symmetry remains unbroken in this regime (cf.\ Sec.~\ref{sec:saddle}). Upon increasing
disorder strength, impurity states proliferate inside the band gaps around zero
energy. Eventually, a mean-field transition occurs into a phase with a finite
average density of states and spontaneously broken replica symmetry. Above this
transition, long-ranged correlations are mediated by the Goldstone modes of the
broken symmetry.

\subsection{Nonlinear sigma-model}

The symmetry-broken phase is described by introducing a bosonic
Hubbard-Stratonovich~\cite{Hubbard1959,Altland2023} matrix field $Q(x)$ as a function of $x$ that decouples the disorder-induced
quartic terms in the replicated action. Reflecting its bosonic nature, this field inherits replica rotation symmetry in the adjoint representation,
$Q(x)\mapsto T Q(x) T^{-1}$. A stationary-phase analysis identifies $Q=\tau_2$ (the Pauli-Y matrix in replica space) as a
representative solution consistent with the antisymmetry constraints of the field.

Allowing the symmetry transformations $T,T^{-1}$ to become spatially dependent,
being reflected by
\[
Q(x)=T(x)\,\tau_2\,T(x)^{-1},
\]
leads to a manifold of Goldstone-mode configurations. The collective dynamics
of these slow modes admits a universal continuum description 
in terms of an effective field theory known as the nonlinear $\sigma$-model~\cite{Wegner1979,Efetbook,Altland2023},
\begin{equation}
\label{eq:Scont}
S[Q]
= \frac{N}{2}\bigg(
g \int_x \tr(\partial_\mu Q\,\partial_\mu Q)
+ \frac{\vartheta}{16\pi} \int_x \mathcal{L}_{\mathrm{top}}
\bigg)\,,
\end{equation}
with
\[
\mathcal{L}_{\mathrm{top}}
= \epsilon_{\mu\nu}\,\tr\!\left(Q\,\partial_\mu Q\,\partial_\nu Q\right).
\]
The integral $\int_x \ev \int \dd^2 x$ is over two spatial dimensions, and $\mu,\nu =1,2$ are the corresponding spatial directions.

The first term penalizes spatial variations of $Q$ and defines a stiffness $gN$ (also interpreted as thermal conductance). Microscopically, the bare value of $g$ depends on the disorder strength $W$ and on details of the band structure near zero energy through the disorder-generated density of states. Explicit expressions for $g(W,a)$ are
obtained from the stationary-phase solution and subsequent gradient expansion;
see Eq.~\eqref{eq:bare-couplings} and
Appendices~\ref{app:sp} and~\ref{app:GradientExpansion}.

The second term is of topological origin. Its integral counts the number of times the field configuration $Q(x)$ as a function of $x$ wraps around its target
manifold. This integer-valued invariant is the \emph{winding number}
$\mathcal{W}\in\mathbb{Z}$, and the corresponding contribution to the action is
$S_{\mathrm{top}} \propto \vartheta\,\mathcal{W}$. Similarly to the stiffness $g$, the bare
value of the topological angle $\vartheta$ depends on disorder strength and band
structure, again given explicitly in Eq.~\eqref{eq:bare-couplings}.
Actions of this form are well known from the theory of the quantum Hall effect~\cite{Pruisken1984,HuckesteinReview1995,MirlinReview}. The
central result here is their emergence as an effective continuum description of
disordered matchgate tensor-network ensembles.

 \begin{figure}[ht]
 \centering
 \includegraphics[width=0.7\textwidth]{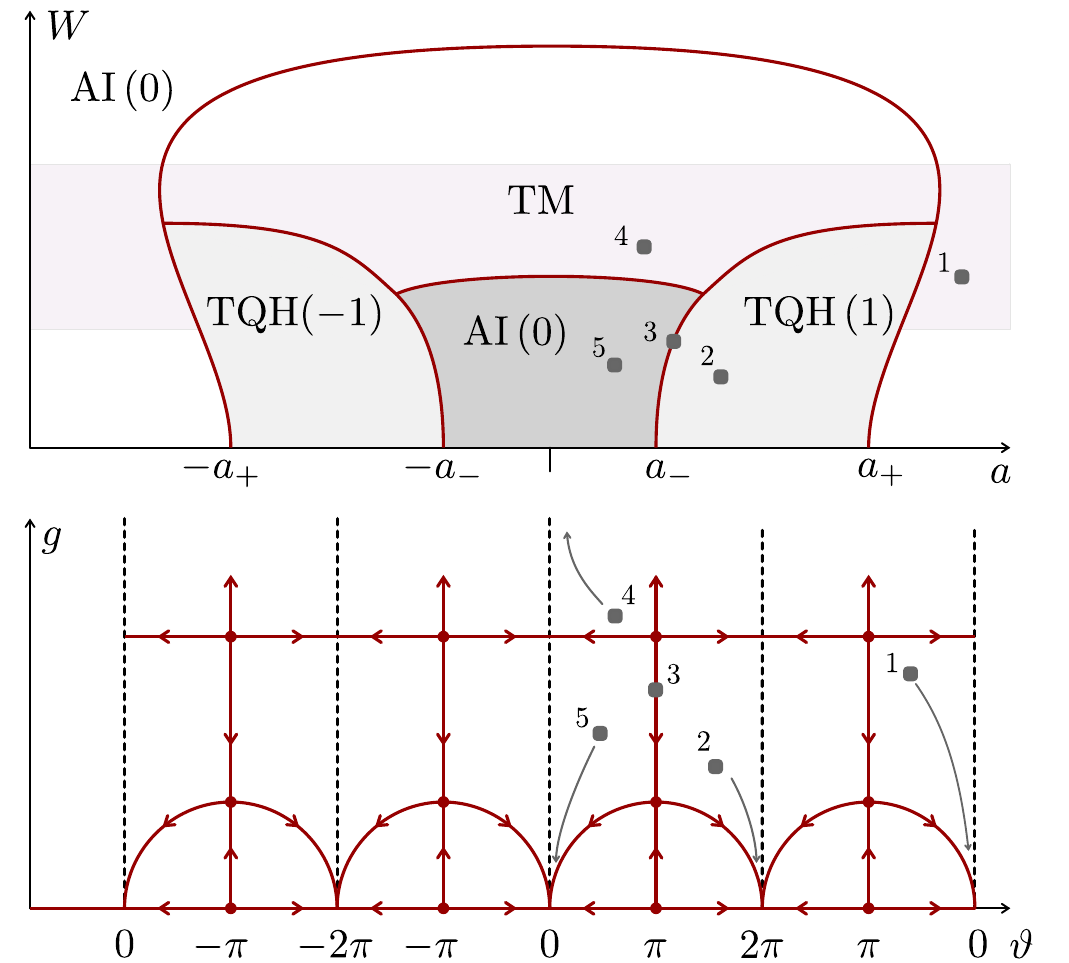}
 \caption{Top: Phase diagram (cf.\ Ref.~\cite{Wang2021}) of the disordered
 class D superconductor, spanned by disorder strength, $W$, and the parameter $a$
 controlling the system's band structure.  The numbers in
 parentheses are the Chern numbers of the different topological phases, \emph{Anderson
 insulator} (AI), \emph{thermal quantum Hall} (TQH).  (The \emph{thermal metal} 
 (TM) phase is
 fundamentally non-topological.) The shaded region marks the parameter range for which our effective theory Eq.~\eqref{eq:Scont} is applicable.  Bottom: schematic flow diagram of the coupling
 constants, $g$ and $\vartheta$ of the field theory. Different parameter
 values $(a,W)$ `initialize' the microscopic values of these constants. Under
 renormalization, i.e., successive  integration over field fluctuations, they 
 flow to qualitatively different fixed points which correspond to
 localization, $(g,\vartheta)=(0,2\pi n)$
 $(1,2,5)$, quantum Hall criticality $(g,\vartheta)=(g^\ast,\pi)$, with a critical
 conductance $g^\ast$, or metallicity $(\infty,0)$	 $(4)$. }
 \label{fig:phase}
 \end{figure}

\subsection{Thermal metal, insulator, and phase diagram}
The effective field theory above is meant to describe physics on length scales much larger than the lattice spacing. A standard way to access such long-distance behavior is to systematically integrate out short-distance fluctuations (coarse-grain) and absorb their effect into scale-dependent effective couplings. In our $\sigma$-model these couplings are the stiffness $g$ (proportional to the thermal conductance) and the topological angle $\vartheta$ (cf.\ Fig.~\ref{fig:phase}). Their two-parameter flow is summarized schematically in
Fig.~\ref{fig:phase}. Different choices of the microscopic parameters $(a,W)$
initialize different bare values of $(g,\vartheta)$. Depending on these initial values, the coupling constants flow towards different regimes -- localization, quantum Hall criticality, or thermal metallicity at long
distances. The flow diagram is divided into small and large bare stiffness $gN\lesssim\mathcal{O}(1)$.

For small bare stiffness $gN\lesssim\mathcal{O}(1)$, the flow enters a
strong-coupling regime beyond quantitative control. General arguments and
previous work~\cite{merzTwodimensionalRandombondIsing2002,Wang2021} indicate the existence of a critical value $g^\ast$ separating
metallic and insulating behavior. For generic initial values of $\vartheta$, the
flow terminates at insulating fixed points $(g,\vartheta)=(0,2\pi n)$, describing
trivial (Anderson) or thermal quantum Hall insulators. These phases are separated by
critical lines at $\vartheta=(2n+1)\pi$, corresponding to thermal quantum Hall
transitions with finite critical stiffness $g^\ast$.

Our effective theory captures the opposite regime of \emph{large} stiffness. When the \emph{bare} stiffness is already large, $gN\gtrsim\mathcal{O}(1)$, the analysis is particularly simple. In that case, field configurations with non-zero winding number $\mathcal W$ are strongly suppressed; thus, the long-distance behavior is dominated by the topologically trivial sector $\mathcal{W}=0$.
In this regime one can track how the stiffness changes under coarse-graining (`renormalization'); to leading order one finds~\cite{Bocquet2000a,MirlinReview}
\begin{equation}
\label{eq:RGEquation}
\frac{\dd(gN)}{\dd\ln L}
= 1 + \mathcal{O}((gN)^{-1}) \,.
\end{equation}
The defining feature here is the positive sign: as we move to larger length scales $L$, the effective stiffness $gN$ \emph{increases}, reflecting a characteristic anomaly of class~D~\cite{MirlinReview,Altland2023}. This is the opposite of what happens in most two-dimensional localization problems, where fluctuations typically \emph{reduce} conductance. Consequently, rather than driving the system into the insulating regime, disorder stabilizes a robust conducting phase: the \emph{thermal metal}.

Deep in this regime, where $gN$ has grown large, the theory becomes
asymptotically Gaussian. Writing the spatial fluctuations $T(x)$ in terms of generators, $T=\exp\,(\frac12 X_a(x) \tau_a)$ with $X_a = X_1,X_3$ and expanding the effective action to quadratic order in the generators $X_a$ yields
\begin{equation}
\label{eq:S_pruisken_gen}
S[X]
= N\,\bigg(
- g \int_x \tr(\partial_\mu X_a\,\partial_\mu X_a)
+ \frac{\vartheta}{16\pi}
\int_x \epsilon_{\mu\nu}\epsilon_{ab}
\tr(\partial_\mu X_a\,\partial_\nu X_b)
\bigg)
+ \mathcal{O}(X^4).
\end{equation}
The first term describes long-ranged correlations in the thermal metal. The second
term is a total derivative, vanishing on infinite or boundary-less geometries
but contributing at system boundaries, where it encodes the thermal Hall
response.

Recall that the effective action above is set up to describe an ensemble of random (matchgate) tensor networks; we thus interpret the long range correlations of the thermal metal phase in this context. To derive correlation functions from the action in Eq.~\eqref{eq:Scont}, we introduce non-local source fields $J_{xy}$ that couple to the slow modes $Q$. This allows us to directly obtain the $n$-th moments of ensemble averaged correlation functions $\mathbb E[C(x,y)^n]$ by taking $n$ derivatives with respect to the source fields $J_{xy}$ and then taking a replica limit. Deferring the technical details to the main text, we observe that, on length scales much larger than microscopic, the effective action above predicts the vanishing of the averaged correlation functions $\mathbb E[C(x,y)] \approx 0 $, where $C(x,y)$ is the realization specific two-Majorana correlation. This follows from the fact that (at the mean-field level) terms linear in the non-local sources do not couple to the local fields. 

This prediction is consistent with our understanding of a random tensor network at large length scales. Namely, in any generic setting away from exact criticality, the disorder-free network has a gapped `Hamiltonian' $H$ and exponentially decaying correlations, negligible on length scales much larger than the correlation length. Adding disorder leads to the (probabilistic) existence of in-gap states ---  eigenstates of $H$ close to zero energy --- which lead to long-range correlations. However, these come with random signs and vanish upon averaging over disorder realizations.

The long-range correlations visible in the effective action of the thermal metal above, however, describe \emph{typical} correlations in a given realization. Concretely, they predict the \emph{second moment} of the correlation function, which derives from terms \emph{quadratic} in the non-local source field and can be read off from the action above to be
\begin{align}
\label{eq:FourPointPlane}
  \mathbb{E} \sqd{\mathcal{C}(x,y)^2 } \propto (gN)^{-1} \ln(|x-y|). 
\end{align} 
Equation~\eqref{eq:FourPointPlane} captures the known logarithmic growth of correlations characteristic of free field
theories in two dimensions. Note that this behavior can change if we consider
the network on a different geometry --- see discussion of the hyperbolic plane in Section~\ref{sec:hyperbolic}.

From a tensor network perspective, we understand the emergence of long range correlations as an accumulation of the sign-random terms mentioned above which cancel in the direct average. Notably, the emergence of long-range correlations is tied to the existence of a non-trivial mean-field solution in the first place. We can thus interpret the disorder-induced symmetry-breaking transition as a transition in \emph{typicality}, meaning that beyond a critical disorder strength the probability to encounter correlations between distant points in any given realization of the tensor network acquires a finite, non-zero value.

\subsection{Nonlinearly deformed matchgate tensor network}

To explore how the physics of our system changes beyond the Gaussian matchgate
limit, we consider nonlinear deformations obtained by adding terms
quartic in the Majorana operators $\theta$. Concretely, we focus on ring-exchange
operators of the form Eq.~\eqref{eq:QuarticTheta}, where the lower indices
$1,\dots,4$ label the sites of the plaquettes shown in
Fig.~\ref{fig:TNLattice}. (For 
a discussion of the physical motivation and
significance 
of this interaction, we refer to Ref.~\cite{WilleQuarticTN:2025}.)

Upon replication, these quartic terms take 
the form
$\sum_r \theta^r_{1}\theta^r_{2}\theta^r_{3}\theta^r_{4}$. Unlike the terms generated by averaging over Gaussian disorder, these are not invariant under continuous replica rotations. As a result,
the continuous replica rotation symmetry present in the free disordered theory
is explicitly reduced to the discrete subgroup of replica \emph{permutations},
which is trivially preserved by the sum over replicas. This reduction from a
continuous to a discrete symmetry represents a qualitative change in the system.

As in the Gaussian theory, the choice of a particular mean-field configuration,
for instance $Q=\tau_2$, spontaneously breaks the remaining replica permutation
symmetry. However, in contrast to the free disordered case, this symmetry
breaking no longer gives rise to continuous Goldstone modes. Instead, the would-be
soft modes acquire a finite mass (equivalently, a finite correlation length).

Quantitatively, the damping of Goldstone modes in the presence of weak interactions
is described by the operator in Eq.~\eqref{eq:InteractionReplicaProjectors}, with a
coupling constant set by the microscopic interaction strength. Considering small
fluctuations $T=\exp(X)$ around the stationary configuration $Q=\tau_2$, this
operator reduces to a mass term, Eq.~\eqref{eq:InteractionGeneratorExpansion},
making the exponential suppression of long-wavelength continuum fluctuations
explicit. At the same time, discrete permutation matrices $T$ remain exact
symmetries of the theory, reflecting the residual replica symmetry pattern.

The qualitative consequences of these two distinct symmetry-breaking scenarios 
--- continuous symmetry breaking with Goldstone modes in the free disordered
theory versus discrete symmetry breaking with massive modes in the interacting
theory --- are substantial. Their implications for correlation functions and the
stability of diffusive behavior in disordered tensor networks will be explored
in detail in forthcoming work.


\section{Gaussian fermionic tensor networks} \label{sec:GFTN}

In this section, we start our construction of the field theory with a brief
discussion of fermionic Gaussian tensor networks in terms
of their Grassmann integral description. For a more detailed discussion we refer
to Ref.~\cite{wille2023topodual}.

\subsection{Fermionic tensor networks} \label{sec:FTN}
In this work, we will consider fermionic tensor networks \cite{PhysRevA.80.042333,MERAF1,PhysRevB.90.085140,CorbozPEPSFermions,PhysRevB.95.245127,FermionicTensorNetworks} with bond dimension two. A tensor $T_\textrm{f}$ in this category with $n$ indices can be conveniently formulated in terms of Grassmann (anticommuting) variables $\theta_j$ as
\begin{equation} T_\text{f} = T_{i_1 \ldots i_n} \theta_{1}^{i_1} \ldots
\theta_{n}^{i_n} \;.
\end{equation}
Note that due to the anticommutativity of the Grassmann variables, the ordering of the $\theta_i$ is an integral component of the definition of the tensor. 

The contraction of fermionic tensors is represented by a Grassmann integral over the two Grassmann variables $\theta_i$, $\theta_j$ that are associated with the two indices to be contracted via
\begin{equation} 	\int \dd \theta_{i} \dd \theta_{j}\, \e^{\theta_{i} \theta_{j}} \ldots \;,
\end{equation}
where '$\ldots$' stands for a collection of fermionic tensors. Note that, again, the ordering of $\theta_i, \theta_j$ in the `integration measure' is part of the definition of the contraction. A fully contracted tensor network can then be represented by the integral
\begin{equation}
\text{TN}_{(C,T)}=	\int (\dd \theta)_C \,\e^{\frac 1 2  \theta^\T C  \theta} \; T^1  \ldots T^{N_s} \;, \label{eq:contracted}
\end{equation}
where $T^1,\ldots,  T^{N_s}$ represent a collection of $N_s$ fermionic tensors, $(\dd \theta)_C$ is a shorthand for the product of all ordered pairs of Grassmann variables to be integrated over and the antisymmetric matrix $C$ represents the (signed) adjacency matrix of the tensor-network. More precisely, the matrix $C$ has a non-zero entry $C_{x i,y j}=1$, if and only if the $i$-th index of a tensor at position $x$ is contracted with the $j$-th index of a tensor at position $y$.

In the following, we will only work with tensors that have even fermion
parity, meaning that $T_{i_1\ldots i_n}=0$ when $i_1+\ldots + i_n$ is odd. In this
setting, the diagrammatic notation of tensor networks remains valid once the
individual ordering of fermionic modes per tensor and the directions of the all
contractions are specified. Specifically, we will consider 
tensor networks on a 2D square lattice where we fix the ordering of fermionic
modes per tensor and the orientation of contracted bonds as shown in
Fig.~\ref{fig:TNLattice}.

\subsection{Fermionic Gaussian tensors} \label{sec:FGT}

We now turn to fermionic Gaussian tensor networks
that correspond to matchgates in the spin  or qubit picture. Fermionic Gaussian tensors capture non-interacting models in condensed matter physics and can be computationally efficiently contracted 
even when defined on non-planar graphs
\cite{Bravyi2009,Jahn,HolographicReview}, rendering them a physically particularly important and conceptually highly relevant family of tensors.

We restrict our attention to tensors of even fermion parity that also obey the normalization condition $T_{0\ldots 0}=1$. In this case, the following definition holds: A (fermion parity even, normalized) tensor $T$ with $n$ indices is a \emph{fermionic Gaussian} (fG) tensor, if and only if there exists a real antisymmetric $n \times n$-matrix $A=-A^\T$ called the characteristic function such that
\begin{equation}
	T_\text{fG} = \e^{\frac 1 2  \theta^\T A  \theta} \;,\quad  \theta^\T=(\theta_1,\ldots,\theta_n)\;. \label{eq:fG}
\end{equation}
It is obvious that the tensor product of two fG tensors $T_1$, $T_2$ with fermionic modes $\theta$, $\xi$ is again a Gaussian tensor
given by
\begin{equation}
	T_\text{1}  T_\text{2} = \e^{\frac 1 2  \Theta^\T (A_1 \oplus A_2)  \Theta} \;, \quad  \Theta=( \theta, \xi)\;,
\end{equation}
and one can show \cite{Bravyi2009} that also the contraction of two fG tensors
is an fG tensor. 

Due to the exponential representation of the individual tensors, we can represent a fully contracted
fermionic Gaussian tensor network as a Grassmann integral that resembles a partition sum of a free fermionic Hamiltonian, namely as
\begin{equation}
	\label{eq:TfGContraction}
	\text{TN}_{(C,A)}=\int (\dd \theta)_C \,\e^{\frac 1 2  \theta^\T (A+C)  \theta} \;, 
\end{equation}
where $A=\bigoplus_i A_i$ is the direct sum of all individual characteristic
functions of the tensors $T_i$ and $\theta$ is the collection of all fermionic modes in the network.

\subsection{Connection to condensed matter path integrals} \label{sec:connect_condmat}

To make the connection to condensed matter systems more apparent, we refer to the
contracted TN as a partition sum $Z$ and introduce the `action' $S$ and
Hamiltonian $H$ as
\begin{equation}
	\text{TN}\to  Z=\int \dd \theta \, \e^{-S}\;, \quad 
	S= - \frac {\i} 2  \theta^\T  H \theta \;, \quad 
	H \coloneqq - \i (A+C)\;. 
	\label{eq:HCA}
\end{equation}
The Hamiltonian $H$ is Hermitian provided our tensors are real, which we will
assume in the following.

From the perspective of many-body physics, the bilinear $\theta^\T H \theta$ is
the Grassmann representation of a system with a free-fermion Hamiltonian $H$.
Such Hamiltonians can be classified into one of ten symmetry classes distinguished
by their antiunittary symmetries~\cite{AltlandZirnbauer}. Besides being Hermitian,
$H=H^\dagger$, our Hamiltonian is antisymmetric by construction,
$H=-H^\T$, identifying it as a member of class~D.

To make this assignment more concrete, consider the second-quantized fermion Hamiltonian
\begin{align}
	\label{eq:Bogolubov-deGennes}
	\hat H
	= c^\dagger h c
	+ c^\dagger \Delta c^\dagger
	+ c \Delta^\dagger c
	= C^\dagger 
	\underbrace{\begin{pmatrix}
		h & \Delta \\
		\Delta^\dagger & -h^\T
	\end{pmatrix}}_{H_{\mathrm{BdG}}}
	C,
\end{align}
where $C=(c,c^\dagger)^\T$ is a Nambu spinor, $h$ is Hermitian, and
$\Delta$ is antisymmetric. This structure describes free-fermion systems of
the lowest degree of symmetry: no time-reversal symmetry, no spin-rotation
symmetry, and in particular no particle-number conservation. The particle-number
symmetry breaking terms proportional to $\Delta$ find physical realizations as
superconducting order parameters, hence the identification of the `least
symmetric' class~D with superconductor physics.

In a path-integral framework, this bilinear form is represented as
$\bar\Psi (E -H_{\mathrm{BdG}})\Psi$,
where $E$ is an energy parameter infinitesimally shifted into the upper
half of the complex plane and $\Psi=(\psi,\bar\psi)^\T$  a Grassmann Nambu spinor representing the complex fermion operator $C$ (cf.\ Ref.~\cite{Altland2023}). Gaussian integration over
$\Psi$ then produces matrix elements of  Green's functions as input for the computation of observables.
Now define the real Majorana variables
\begin{equation}
	\theta_1 \coloneqq 
    \frac{1}{2}(\psi+\bar \psi), \qquad
	\theta_2 \coloneqq - 
    \frac{\i}{2}(\psi - \bar \psi) \,,
\end{equation}
and $\theta=(\theta_1,\theta_2)^\T$. It is straightforward to verify that the bilinear form above, evaluated at the
symmetry point of the superconductor spectrum $E=0$\footnote{Recall that the spectrum of
any superconductor is symmetric around $E=0$.}, is unitarily equivalent to $-\theta^\T H \theta$, with the real and antisymmetric matrix
\begin{align}
	\label{eq:BdGVsMajorana}
	H =
	\begin{pmatrix}
		2 \,\mathrm{Im}(h-\Delta)  &
		2 \, \mathrm{Re}(\Delta - h)  \\
		2 \, \mathrm{Re}(\Delta + h) &
		2 \,\mathrm{Im}(h+\Delta)
	\end{pmatrix} \;.
\end{align}

We conclude that the tensor-network bilinear form $- \theta^\T H \theta$ of our
tensor network  affords an interpretation as
the path-integral representative of a class~D superconductor action at the
symmetry point of the spectrum. Note that any (even-dimensional) antisymmetric
matrix can be represented in the form \eqref{eq:BdGVsMajorana}, i.e., 
the bridge
to superconductor physics is not restricted to particular forms of correlation
matrices, provided they are even-dimensional\footnote{Odd-dimensional theories correspond to condensed
matter systems harboring local isolated Majorana modes. However, we will not
discuss this ramification any further.}. This analogy will frequently be useful in
establishing parallels between the physics of the tensor network and structures
present in superconducting contexts. 

This being said, it is important to have in mind that the tensor network field
integral remains fundamentally different from the coherent state path integrals
employed in many-body physics. The latter are Trotterizations of real or
imaginary time evolution operators $\exp(\tau H)$, with resolutions of unity
between individual time slices realized through coherent states. On the level of
effective actions, this leads to structures of the form $\int_\tau \dd t (\bar \Psi
(\partial_\tau + E-H)\Psi)$, where the time derivative accounts for the
dynamical coupling between neighboring slices. In contrast, our tensor network	
represents a single quantum state (supported on its non-contracted indices),
without such dynamical structure. The analogies to superconductor physics
mentioned above notwithstanding, this is a major difference. For example,
nonlinear terms featuring in dynamical path integrals, i.e., genuine particle
interactions, are much harder to get under control than the quartic terms
discussed later in this work.

\subsection{Expectation values and correlation functions} \label{sec:corr}
The framework of fermionic Gaussian tensor networks allows for a rather straightforward calculation of correlation functions.
An arbitrary correlation function of fermionic Gaussian tensors can be expressed as a sum of elementary correlation functions of the form
\begin{equation}
    \label{eq:CorrelationFunction}
	\mathcal C(\theta_1,\theta_2,\ldots) \coloneqq \frac 1 Z \int \dd \theta \, 
    (\theta_1 \theta_2 \ldots )\,\e^{\frac \i 2  \theta^\T H  \theta}.
\end{equation}
Such objects can be evaluated by the rules of Gaussian Grassmann integration
{(see, e.g., Refs.~\cite{Altland2023,BarrySimon})}, reflecting a contraction of
a matchgate tensor network \cite{Bravyi2009, Jahn}. For instance, for the
simplest case of a two-point correlation function, it holds
\begin{equation} \label{eq:Cxy}
    \mathcal C(\theta_i,\theta_j) = \i H^{-1}_{ij} \,. 
\end{equation}
By the same principle correlation functions of higher (even) order in Grassmann
insertions yield sums of products of two-point functions, in accordance with Wick's theorem.

\begin{figure}
\centering
\includegraphics[width=0.3\textwidth]{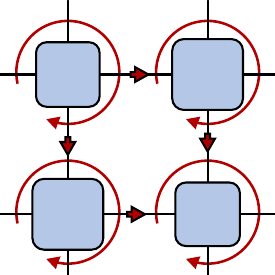}\hspace{1.5cm}
\includegraphics[width=0.3\textwidth]{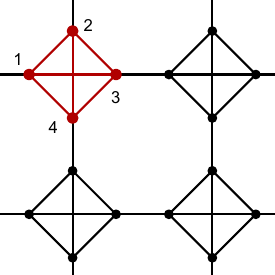}
\caption{Left: Fermionic tensor network on a square lattice showing the ordering of fermionic modes (clockwise red arrow) per tensor and the bond orientations (red arrows). Right: Representation of the pseudo Hamiltonian corresponding to the tensor network on the left in case the tensors are Gaussian, cf.\ Eq.~\eqref{eq:HCA}. The labeling of modes in the unit-cell (red) is inferred from the ordering of the fermionic modes per tensor.}
\label{fig:TNLattice}	
\end{figure}

\subsection{Minimal Hamiltonian: Haldane-Chern insulator} \label{sec:haldane-chern}
In this section so far, our discussion of fermionic Gaussian tensors has applied to general realizations of network geometry, captured by the connectivity matrix $C$, and coupling strength between the fermionic modes $\theta$, captured by the matrix $A$. 
For concreteness, we now consider the minimal non-trivial Hamiltonian\footnote{From here on, we will refer to the kernel $H$ of the tensor network bilinear form $\theta^\T H \theta$ as the `Hamiltonian', exploiting the analogy to fermionic systems discussed in Sec.~\ref{sec:connect_condmat}.}
\begin{equation}
	H_0(a) = - \i \rnd{C_\text{sq} + \textstyle{\bigoplus_x A(a)}} \;, \label{eq:HC}
\end{equation}
where $C_\text{sq}$ is the (signed) adjacency matrix of the oriented square lattice depicted in Fig.~\ref{fig:TNLattice}, and the matrix $A$ is defined by a single parameter $a$ via $A_{\alpha \beta }(a)=a\,\textrm{sgn}(\beta-\alpha)$. In order to unambiguously define the network, we also need to choose an ordering of the four fermionic modes per tensor and an orientation of the bonds stored in the adjacency matrix $C$. Figure~\ref{fig:TNLattice} illustrates our choice. 

\begin{figure}
	\centering
	\includegraphics[width=0.32\textwidth]{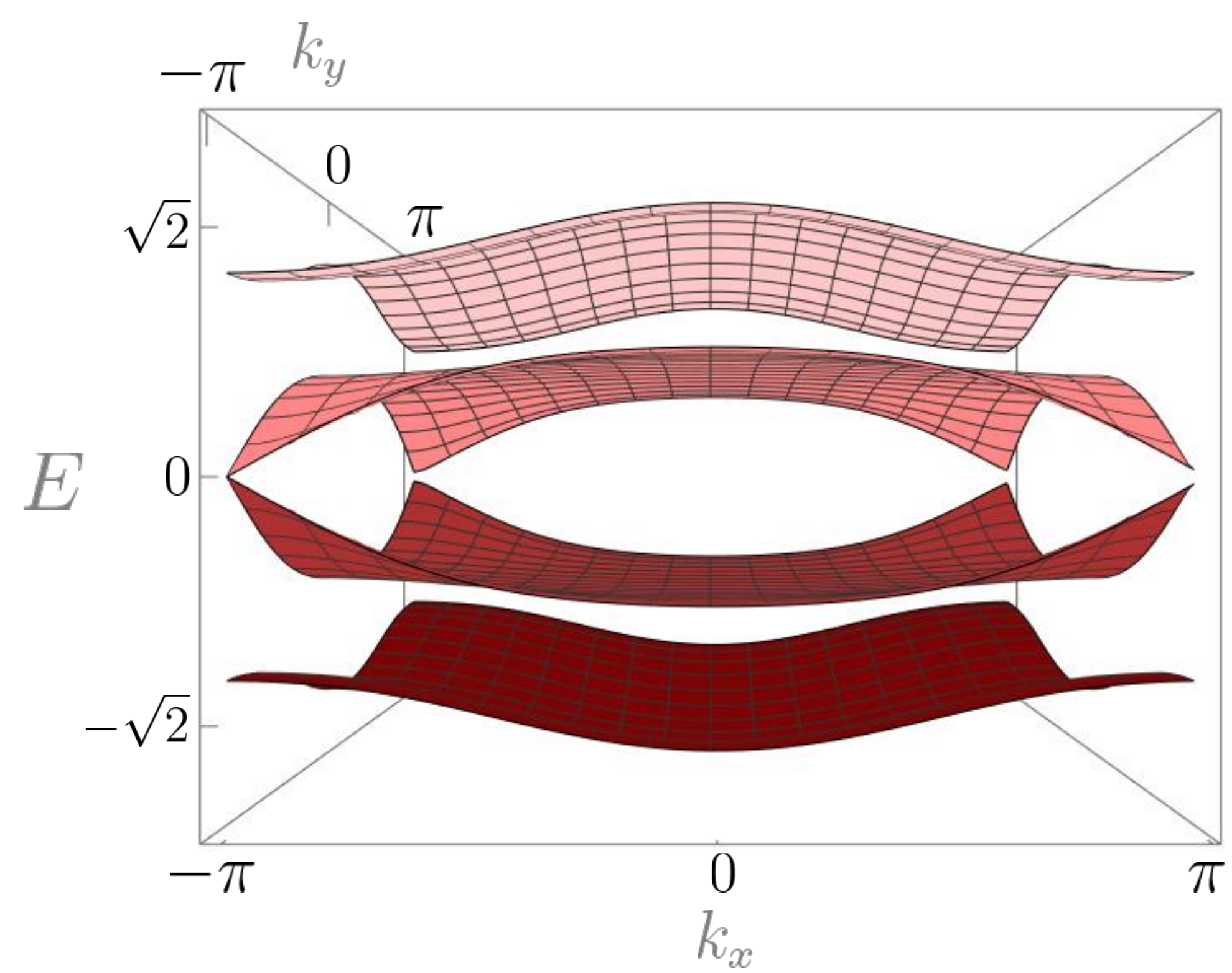}
	\hfill
	\includegraphics[width=0.32\textwidth]{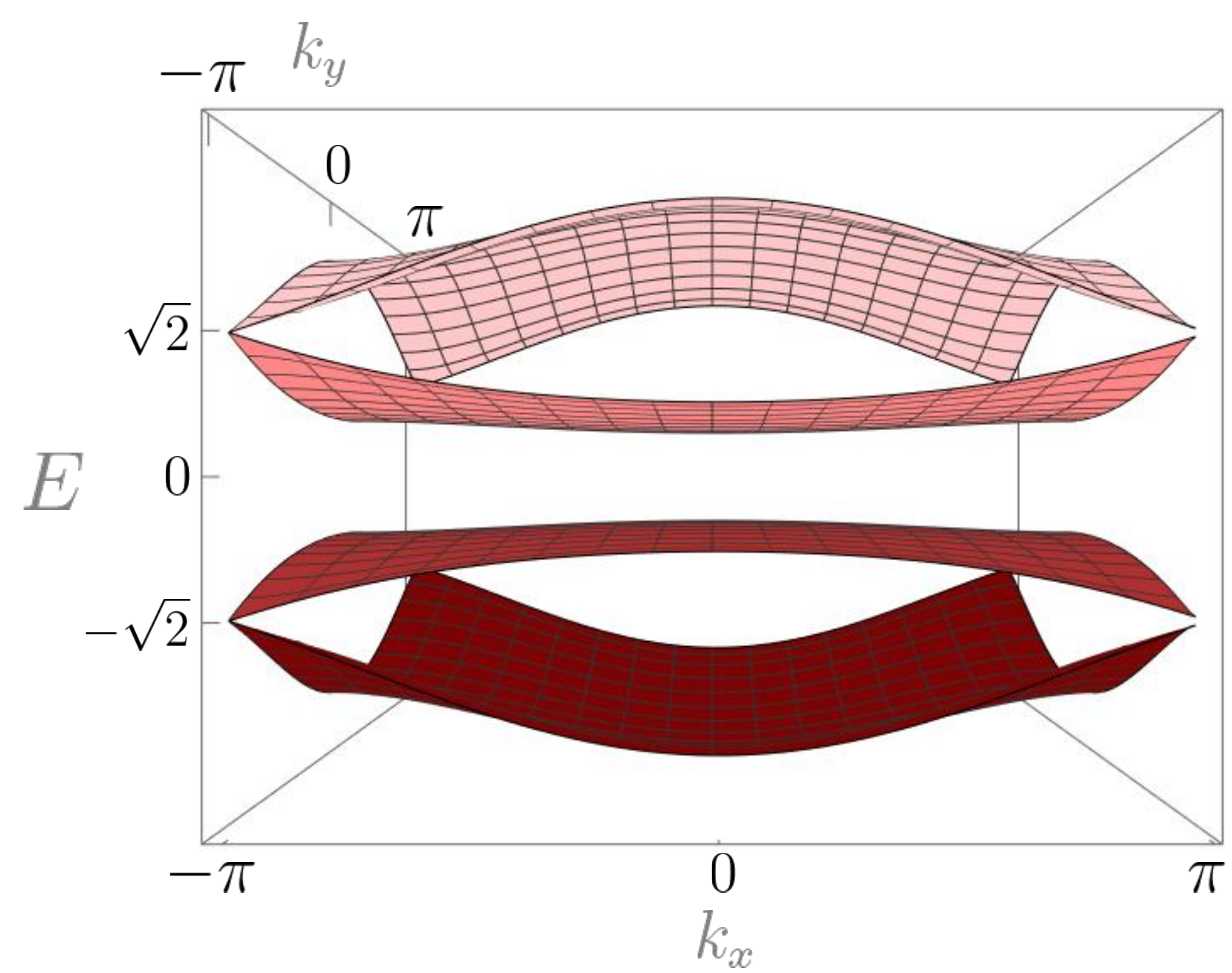}
	\hfill
	\includegraphics[width=0.32\textwidth]{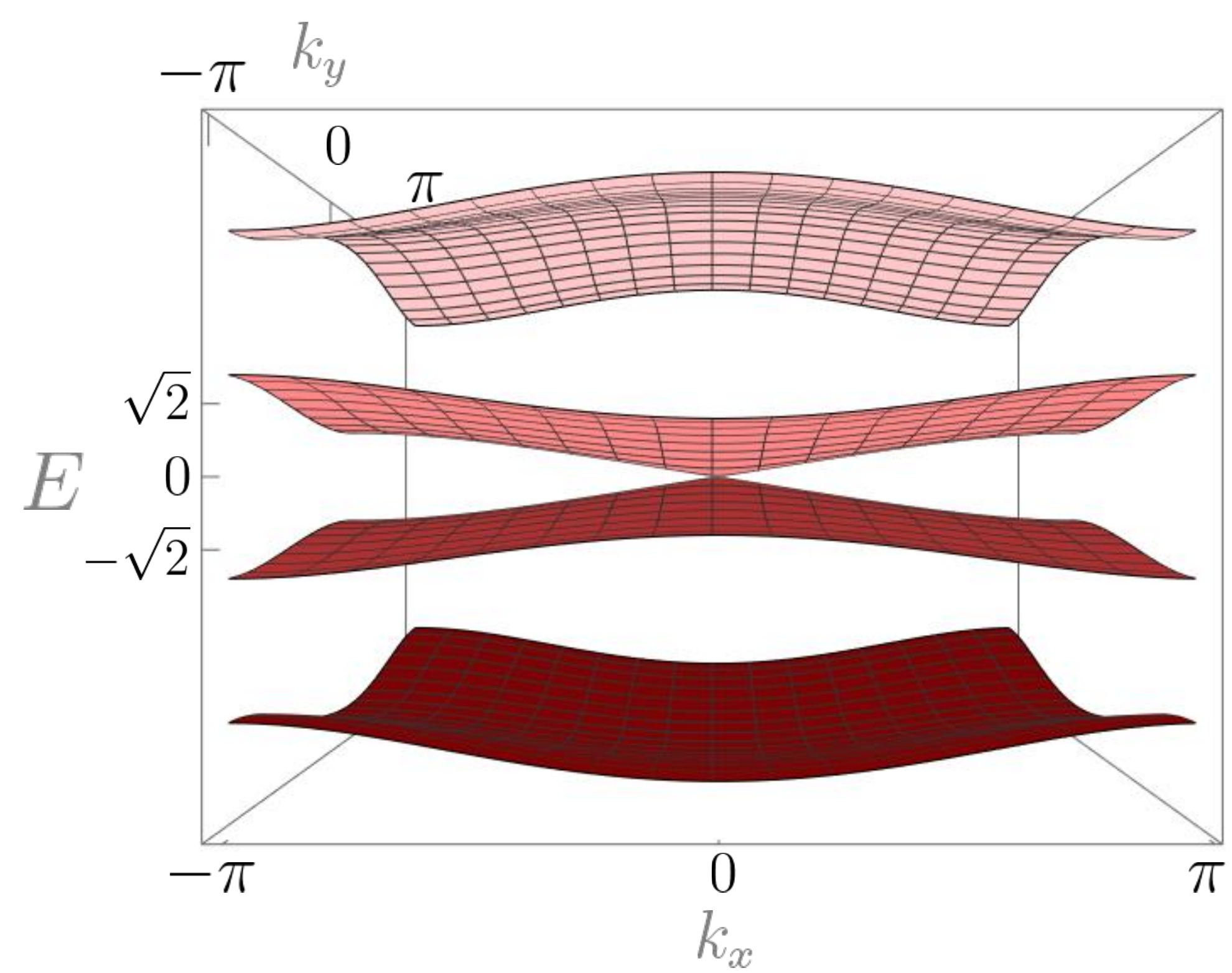}
	\caption{Band structure of the (four-band) pseudo-Hamiltonian $H= -\i (C +\bigoplus A)$ for $A_{\alpha \beta}=a \operatorname{sgn}(\beta-\alpha)$ at $a=a_-$ (left), $a=1$ (center) and $a=a_+$ (right).}
	\label{fig:bands}
\end{figure}

Since the system is translationally invariant, it is more convenient to evaluate the bilinear form $\theta^\T H \theta$ in momentum space, leading to the compact form $\sum_{q} \theta^\T_{-q} H(q) \theta_{q}$ (cf.\ Appendix~\ref{app:FT} for details). Here, $H(q)$ is now a $(4
\!\times\! 4)$-matrix defining a \textit{four-band} Hamiltonian in the parlance of condensed matter theory. Only two of these bands are independent, because the antisymmetry of $H=-H^\T$ implies the symmetry of its spectrum around $0$ (in other words, the eigenvalues come in pairs $\pm \eps$).     
For generic values of $a$, this four-band Hamiltonian is gapped. However, at $a=a_\pm \coloneqq \sqrt{2} \pm 1$, we find a gap closing at $E=0$ for $k_-=(k_1,k_2)=(\pi,\pi)$ and $k_+=(0,0)$, respectively. The band structure of the four-band Hamiltonian is shown in Fig.~\ref{fig:bands}.

Considering the two-point correlation function in Eq.~\eqref{eq:Cxy} in the eigenbasis of $H$, it is  apparent that the
behavior of two-point functions is dominated by the smallest
eigenvalues. Therefore, a projection onto the subspace of eigenstates of $H(q)$ with eigenvalues closest to zero is a justified approximation, provided we stay close enough to the critical points at $a_\pm$. As explained in more detail in Ref.~\cite{wille2023topodual}, this so-called \emph{two-band approximation} leads to 
\begin{equation} 
\spl{
 	\label{eq:Haldane_Ham}
	H_\pm^{(2)}(q) &= \sum_{a=1}^3 h_a(q)\sigma_a \,, \quad \text{with} \\
	h_1 &= \frac 1 2 \sin q_1 \,,\quad 
	h_{2,\pm} = \frac 1 2 (2+m_\pm-\cos q_1 -\cos q_2)\,,\quad  
	h_3= \frac 1 2  \sin q_2 \,,
}
\end{equation}   
where $q_\mu=k_\mu-k_{\pm,i}$ are the momenta relative to the  gap closing points and $m_\pm= \pm 2(a-a_\pm)$ is the width of the energy gap at $q_1=0=q_2$. 

The operator $H^{(2)}$ is otherwise known as the Hamiltonian of a two-dimensional \emph{Haldane-Chern (HC) insulator} \cite{Haldane1988,QWZ2006}, appearing, 
e.g., in the context of quantum anomalous Hall effect \cite{Moreno2023}. Depending on the sign of the gap parameter $m$, it is found in one of two topologically distinct phases characterized by their Chern numbers. 
For $m>0$ (i.e., $a_- < a < a_+$), the Chern number is $C=1$, corresponding to the topological phase. For $m<0$ (i.e., $a < a_-$ and $a > a_+$), the Chern number is $C=0$, corresponding to the trivial phase. At the critical point $m=0$, the spectral (eigenvalue) gap between the two bands closes. For all other values of $m$, the band gap remains open, indicating that we are dealing with an insulator\footnote{Note that the two-band approximations above are valid only around $a_\pm$. In particular, at $a=1$, there is an additional band crossing at $E=1$ of the two lowest and two highest bands. Hence, $H_+$ is not 'adiabatically connected' to $H_-$.}.

We thus conclude that, for the minimal choice of the tensor network Hamiltonian as in Eq.~\eqref{eq:HC}, the low-energy physics of the model is accurately captured by the Haldane-Chern Hamiltonian in Eq.~\eqref{eq:Haldane_Ham}. In the following,
this two-band model is going to be the staple of our further analysis of fermionic Gaussian tensor networks, with randomness included.

\paragraph{Weakly disordered Dirac fermions.}
In order to develop a first intuition for the effects of disorder, we can compare our system to weakly disordered Dirac fermions. The correspondence is established by considering the vicinity of a phase transition point between the two
different insulator phases above and observing that random
fluctuations --- e.g., in the parameter $a$ --- manifest  themselves in fluctuations of the
coefficient $m$. Focusing on the lowest lying states of the eigenvalue spectrum, we linearize $\sin(q_\mu)\mapsto q_\mu$, $\cos (q_\mu)\mapsto 1$ and (after a change of basis $\sigma_2 \leftrightarrow \sigma_3$) arrive at the Dirac Hamiltonian
\begin{align} \label{eq:H_Dirac}
   H_\text{Dirac}=  q_1 \sigma_1 + q_2 \sigma_2 + m \sigma_3 \;, 
\end{align}
where we dropped an overall factor $1/2$, and $m$ is continuum notation for a
randomly fluctuating `mass' coefficient. For $\langle m \rangle=0$, we
describe a situation at criticality, where the clean band gap closes. 

Random mass Dirac fermions have been thoroughly investigated in the
literature \cite{TAKEDA1999randomMass,Bocquet2000a,Pan2021randomMass}. A \emph{renormalization group} analysis shows that weak mass perturbations
are (marginally) irrelevant, i.e., the system behaves effectively clean when
viewed at large length scales. In this regime, the physics of two insulating
phases separated by a topological quantum critical point remains largely
unaltered. However, a more sophisticated extension of the RG approach to two-loop order
shows the presence of a critical disorder concentration, i.e., the variance
$\langle m(x)^2 \rangle$ exceeding a critical value, above which the system 
is in a symmetry broken phase characterized by the non-vanishing average density of states at zero energy~\cite{Bocquet2000a,MirlinReview}. The
RG analysis identifies it with the thermal metal phase. One should not be
confused by the seemingly paradoxical situation that increasing disorder leads
to metallicity: the clean Dirac system does not support states at zero energy.
Even at criticality, it is a `semi-metal' halfway between metal and insulator.
Disorder makes it more `metallic' by adding states which may participate in the
buildup of correlations and transport.


\section{Ensemble average and nonlinear sigma-model}
\label{sec:ft-derivation}

In the previous section, we have (for the most part) discussed the construction of fermionic Gaussian tensor networks with their couplings being fully \emph{deterministic} (non-random). We refer to this as the clean limit.
In this section, we go beyond the clean limit and construct an
effective \textit{field theory} describing ensemble averages of random matchgate tensor networks.

\subsection{Field theory construction}

For the derivation of the field theory we employ the replica trick and trade the
ensemble average over random tensors for an integral over a  matrix-field in
replica space. We evaluate the resulting theory with stationary phase methods
and arrive at an effective soft mode action in the family of nonlinear
$\sigma$-models\cite{Efetbook,Altland2023}. These steps depend solely on the
(class~D) symmetry of the system and on the presence of a large number of bond
degrees of freedom. Yet, the subsequent derivation of an effective continuum
theory by gradient expansion makes an extensive reference to the properties of
the underlying tensor network: in the presence of disorder, the topological
transition between phases with different Chern numbers becomes a quantum Hall
transition in the universality class of the thermal quantum Hall
effect~\cite{choCriticalityTwodimensionalRandombond1997,merzTwodimensionalRandombondIsing2002}.
In the weak disorder limit, the low-energy effective field theory describing the
transition is a \emph{nonlinear $\sigma$-model} (NLSM) enriched by a topological
term~\cite{Bocquet2000a}.
In the following, we derive the NLSM under the weak disorder assumption and then review
what is known about its behavior at large distance scales, where even nominally weak disorder strongly alters the physics of the clean system.

\subsection{Ensemble average} \label{sec:replica} 
Shifting our focus away from the microscopic details of a single tensor network, we now ask instead what can be said about its \textit{universal} properties which are robust under variations of the system's parameters. To this end, we replace a single tensor network by an ensemble of networks with fluctuating parameters. In practical terms, this corresponds to adding \emph{disorder} to a clean system and averaging over different disorder realizations. 

Physically interesting theories emerge in cases where the spectrum of the clean system does not contain a large gap around zero `energy'. In the opposite scenario, weak disorder will only lead to a slight deformation of the spectral gap and correlation functions will still be exponentially damped in its size. In our case, we choose to start from the clean system $H_0$ near the critical point (cf.\ Sec.~\ref{sec:haldane-chern}).

\paragraph{Large-$N$ limit.}
To make analytic progress with the derivation of an effective theory, it is mandatory to resort to a \emph{large N} limit of some kind. In our case, a natural large $N$ parameter is introduced by considering $N$ non-interacting copies of the original tensor network or, equivalently, using a single tensor network with tensors of bond dimension $2^N$ ($N$ parallel bonds of dimension two).

Viewed from the physics perspective, the necessity of the large-$N$ limit follows from the following observation in analogous condensed matter systems. At the gap closing transition, the clean Chern insulator turns into a so-called semimetal, whose Dirac-like gap closing point supports an electrical conductance of the order of the conductance quantum, 
that is, $e^2/ h$. In the presence of disorder, systems with a conductance this low are governed by strong quantum fluctuations, which no analytical approach can get under quantitative control. One way to handle the situation is to consider not one but a large number $N$ of coupled Dirac points, enhancing the conductance to values $\sim N$ in units of the conductance quantum. This, then, is a `good' metal that, in the presence of disorder, turns into a quantum conductor amenable to perturbative approaches of quantum field theory. 

With this in mind, we consider the generalized clean action $\theta^\T H_0 \theta \mapsto \theta^\T (H_0 \otimes \id_N) \theta$, with $N$-fold extended fermionic modes $\theta_{x,i}$, where $i=(\alpha,c)$ is a composite index of the unit cell position $\alpha=1,\ldots,4$ and the layer index $c=1,\ldots,N$. Close to criticality, the clean $N$-layer system
describes $N$ isolated gap closing points. We now consider disorder isotropically coupling these modes locally and introduce the ensemble $H=H_0 \otimes \id_N + \tilde H$, where $H_0$ is fixed and $\tilde H \equiv - \i {\tilde A}= - \i \bigoplus_x {\tilde A}_x$ is given by a collection of `random' elements ${\tilde A}_x$ coupling all modes at $x$ with zero mean and second moment proportional to the disorder strength $W$
 \begin{equation} \label{eq:R_moments}
    \av{ {\tilde A}_{x,ij} \; {\tilde A}_{y,kl} }_{{\tilde A}} = \frac{W^2}{4N} \delta_{xy} (\delta_{ik} \delta_{jl} - \delta_{il} \delta_{jk}) \,,
\end{equation}
where the Kronecker deltas follow the multi-index convention $\delta_{ij}=\delta_{(\alpha,c)(\beta,d)}\equiv \delta_{\alpha \beta} \delta_{cd}$.

\begin{figure}
\centering
	\includegraphics[width=0.33\textwidth]{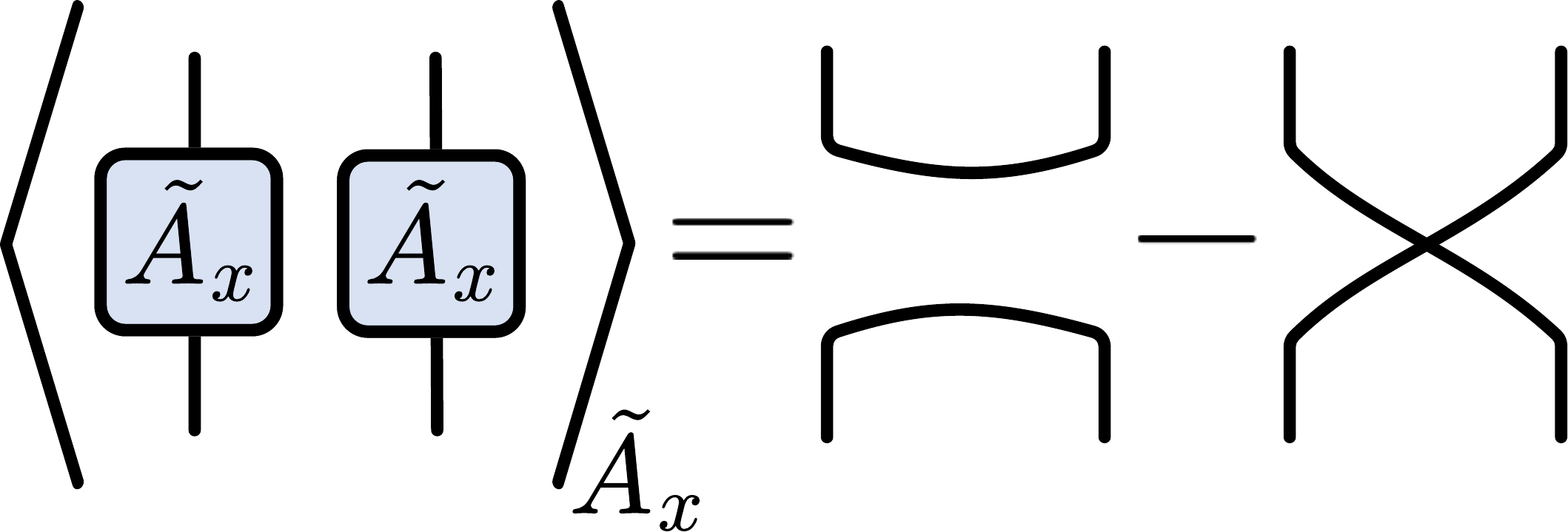}
	\caption{Graphical illustration of the contraction rule in  Eq.~\eqref{eq:R_moments} (up to prefactors).}
	\label{fig:R2}
\end{figure}

\paragraph{Replica partition function.}
In the presence of disorder,  $\tilde A$, tensor network correlation functions such as those in Eq.\ \eqref{eq:Cxy} become statistically distributed. 
The average over such functions, and over higher moments characterizing their statistics, is complicated by the presence of the (realization specific) partition sums $Z$ featuring in the denominator of the definition Eq.~\eqref{eq:CorrelationFunction}. 
There are two established techniques to deal
with this complication ~\cite{Efetbook,Altland2023}: the supersymmetry and the replica method. 
Adopting the latter \footnote{For the quantum information readership, the replica method is basically making use of copies of quantum systems to have access to higher moments.}, we introduce $2R$ copies\footnote{For reasons that will become clear momentarily, the symmetries of the problem motivate working with an even number
of replicas.} (replicas) of the original system, $H\mapsto H \otimes \id_{2R}$. Replicating the Grassmann fields in an analogous way, $\theta = \crl{\theta^r_x}$, $r=1,\dots,2R$,
where we suppress the intra-cell indices $(x,i)\mapsto x$ for brevity, the
moments of correlation functions can now be represented as 
\begin{equation} \label{eq:ReplicaCorrelationFunction}
\begin{split}
	\mathcal{C}(\theta_{x},\theta_{y})^n     
 	&= \lim_{R\to 0} \int \dd \theta \, 
 	\e^{\frac{\i}{2}\theta^\T (H\otimes  \id_{2R}) \theta } \textstyle{\prod}_{r=1}^n	(\theta_{x}^r \theta_{y}^r) 
 	\qquad\quad (n \leq 2R)
 	\\
 	&= \lim_{R\to 0} Z^{2R} (\i H_{xy}^{-1})^{n} = (\i H_{xy}^{-1})^{n} \,,
\end{split}
\end{equation}
with the single-replica partition function $Z$ defined in Eq.~\eqref{eq:HCA}, and the elementary two-point correlator given by Eq.~\eqref{eq:Cxy}. 

To conveniently represent these functional expectation values in our formalism, we introduce a source matrix $J$ acting in TN-index and replica space and define $j^r_{xy}$ as the coefficient of $\theta_x^r\theta_y^r$ in $\theta^{\mathsf T}J\theta$. Concretely, $J$ is antisymmetric in the TN indices and diagonal in replica space; for a fixed pair $(x,y)$ the only non-zero entries are $J^{rr}_{xy}=+\tfrac12 j^r_{xy} = -J^{rr}_{yx}$, so that $\theta^{\mathsf T}J\theta=j^r_{xy}\theta_x^r\theta_y^r$.
With this definition, the ensemble averaged moment of the correlation functions
assumes the form 
\begin{align}
    \label{eq:CorrelationFunctionSource}
   \mathbb{E} \sqd{\mathcal{C}(\theta_{x},\theta_{y})^n }= \lim_{R\to 0} \rnd{2/\i}^n\,
   \del_{j^1} \cdots \del_{j^n}
   \big|_{J=0}     
    \av{ \int \dd \theta \, \e^{\frac{\i}{2}\theta^\T (H\otimes  \id_{2R}+J)\theta } }_{{\tilde A}},
 \end{align} 
where here and throughout $\mathbb{E} [\dots] \coloneqq \av{\dots}_{\tilde{A}}$ denotes averaging over randomness. Note that, in
contrast to the layer indices, 
replicas are decoupled before disorder average; the average then introduces inter-replica interactions. (To proceed, we will keep the number of replicas finite and fixed, $2R \in \Zbb$, and only after deriving the effective long-range theory take $R\to0$ in replica-invariant observables.)

\paragraph{Introducing the matrix field.}
Turning to the ensemble average, we note that the average over ${\tilde A}$ as specified in Eq.~\eqref{eq:R_moments}
factorizes into averages over individual local ${\tilde A}_x$, allowing us to focus on a single one of those. Further, we perform a transformation that exchanges an integral over a matrix $A_x$ in the space of physical indices (and a unit matrix in replica space) for an integral over a matrix $B_x$ in replica indices (but a unit matrix in the space of local modes). The  transformation is implemented by the integral identity
\begin{equation} \label{eq:color_flavor}
\left\langle \exp \left( \frac 1 2 \theta^\T ({\tilde A}_x \otimes \mathbb 1_{2R})\theta \right)\right\rangle_{{\tilde A}_x}=\left\langle \exp \left( \frac 1 2  \theta^\T (\mathbb 1_{4N} \otimes B_x)\theta \right) \right\rangle_{B_x} \;, 
\end{equation}
where $B_x$ are antisymmetric Hermitian $2R\times 2R$-matrices integrated over the Gaussian distribution
\begin{equation}
	\av{\dots}_{B_x} = \int \dd B_x \, (\dots) \, \e^{-\frac{N}{W^2} \tr \rnd{B_x}^2} 
\end{equation}
with the measure $\dd B_x$ normalized as $\int \dd B_x \e^{-\frac{N}{W^2} \tr \rnd{B_x}^2} = 1$.

\begin{figure}
\centering
	\vspace{3mm}
	\includegraphics[width=0.95\textwidth]{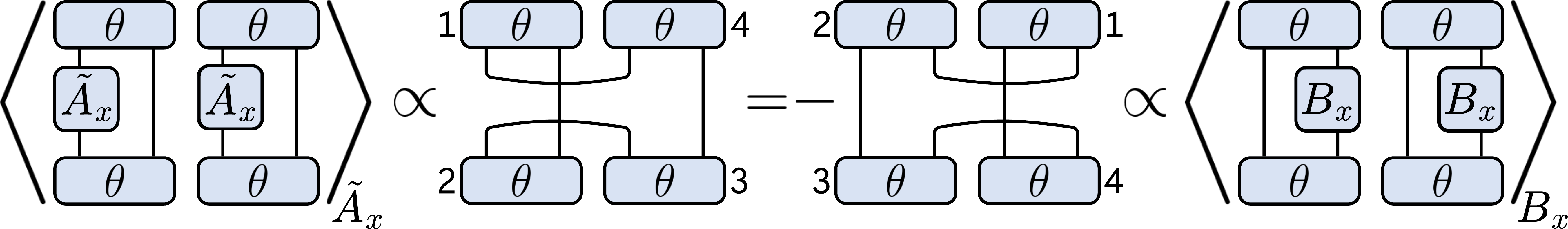}
	\vspace{1mm}
	\caption{Trading averages over random tensors $\tilde A$ for averages over matrix fields $B$. The equality above is derived from the contraction rule for ${\tilde A}_x$ averages in Fig.~\ref{fig:R2}, an analogous contraction rule for $B_x$ averages (not shown) and elementary diagrammatic deformations. }
	\label{fig:CF}
\end{figure}

The statement of Eq.~\eqref{eq:color_flavor} can be derived by expanding the exponentials on both sides
and comparing the averages order by order. As the distributions are Gaussian and
higher moments follow via Wick's theorem, it is sufficient to verify the relation
for the second order term (here and in the following repeated indices are summed over, unless indicated otherwise). Using the contraction rule in
Eq.~\eqref{eq:R_moments} and dropping the 
site label $x$, we find  
\begin{equation*}
	\av{{\theta^r_i {\tilde A}_{ij} \theta^r_{j}} \; {\theta^s_k\, {\tilde A}_{kl} \theta^s_{l}}}_{{\tilde A}} = \frac{W^2}{2N} \theta^r_i \theta^s_i \theta^s_j \theta^r_j \,.
\end{equation*}
The right-hand side has the structure of a matrix-multiplication in replica space. However, under a permutation of the Grassmann variables, it may
just as well be interpreted as a matrix-multiplication in physical space (indices $i,j,...$)
and be represented as an average  
\begin{equation}
\frac{W^2}{2N} \theta^r_i \theta^s_i \theta^s_j \theta^r_j =-\frac{W^2}{2N}  \theta^r_j \theta^r_i \theta^s_i \theta^s_j =\av{{\theta^r_i B^{rt} \theta^{t}_{i}} {\theta^s_j B^{su} \theta^{u}_j}}_{B} 
\end{equation}
over a matrix $B$ drawn from a Gaussian distribution.
The equalities above are represented in pictorial notation in Fig.~\ref{fig:CF}.

Having established the local transformations, we can perform them
at each site and collect the emerging local fields $B_x$ into the
block-diagonal matrix $B\coloneqq \bigoplus_x B_x$. By doing so, we obtain the ensemble averaged partition sum as 
\begin{equation} \label{eq:Z_theta_B}
    \mathcal Z \coloneqq \mathbb{E}\rnd{Z^{2R}} = \int \dd B \, \e^{-\frac{N}{W^2} \tr B^2 }  \int \dd \theta \,\e^{\frac{1}{2} \theta^\T ( B \,+\, \i  H_0) \theta} \,  \,,
\end{equation}
where we have used a shorthand notation  $\id_{4N} \otimes B + \i H_0 \otimes
\id_N \otimes \id_{2R} \coloneqq B + \i H_0$, omitting tensor products with unit
matrices, and temporarily suppressed the source matrix $J$ for notational brevity. We finally note that the passage from a Gaussian distributed $A_r$ to
a likewise Gaussian distributed $B$ is a Hubbard-Stratonovich transformation
frequently applied in cases like the present one, where it is expedient to
introduce a physically motivated effective field variable.    

Since the integral over Grassmann variables in Eq.~\eqref{eq:Z_theta_B} is Gaussian, we can integrate out the fermionic variables as $\int \dd \theta \e^{\frac{1}{2} \theta^\T (B + \i H_0) \theta} = \operatorname{pf} (\i B - H_0)$\footnote{The factor $\i$ multiplying the exponent drops out because the matrix dimension is a multiple of $4$.} .
Re-exponentiating the 
Pfaffian, using that $\operatorname{pf}^2=\det$ and applying the identity $\ln \det = \tr \ln$, we arrive at the averaged replicated partition function
\begin{equation}
    \label{eq:action_B_trln}
    \Zcal = \int \dd B \, \e^{-S[B]}, \qquad
    S[B] =N\rnd{\frac{1}{W^2} \tr\!_\textrm{r,s} B^2 - \frac{1}{2} \tr\!_{\textrm{r,s,o}} \ln \rnd{\i B - H_0}} \,,
\end{equation}
where we have used a notation $\textrm{tr}_{\textrm{r,s,o}}$ indicating whether a trace runs over (r)eplicas, (s)ites, or (o)rbitals. (A trace without subscript extends over all indices.)

Here, the upfront factor of $N$ reflects the isotropy of the
coupling between the $c$-indices: \emph{on average}, the system is uniform in
layer space, and the corresponding trace produces a multiplicative factor.
Conceptually, this principle boosts correlations in the system (corresponding to its   `conductance' in the condensed matter way of reading the situation) by a factor of $N$. Technically, the
factor $N$  will stabilize our subsequent stationary phase analysis.   

Another point that will work to our advantage is that the action in Eq.~\eqref{eq:Z_theta_B} possesses a
continuous symmetry group, $G$: Transformations $\theta \mapsto T \theta$ , $T \in G$ are symmetries of the
microscopic action provided $\theta^\T H \theta \mapsto \theta^\T T^\T H T
\theta=\theta^\T H \theta$, or, equivalently, $T^\T H T = H$. The maximal group of
transformations satisfying this condition is $G= \textrm{O}(2R)$, 
i.e., 
rotations in replica space commutative with $H\otimes \id_{2R}$. On the level of
the effective action in Eq.~\eqref{eq:action_B_trln}, these symmetry transformations are represented as $B\mapsto T B
T^\T$, where the cyclic invariance of the trace implies the invariance of the action.

\subsection{Stationary phase program} \label{sec:saddle}
Due to the strong nonlinearity (the logarithm) of the action
\eqref{eq:action_B_trln}, the integral over matrix fields $B$ cannot be done in
closed form. However, the large factor of $N$ in front of the action implies strong damping of fluctuations away from stationary configurations $B_\ast$ defined by $\delta_B S[B_\ast]=0$. 
In 
the following, we will identify these configurations and observe that some of them \emph{break} the aforementioned $\Ort{2R}$ symmetry. This implies the existence of an entire manifold 
of solutions $B_\ast\mapsto T B_\ast T^{-1}$. The generalization from a uniform $T$ to a \emph{field} $T_x$ with slow site-dependence identifies the soft modes of the problem, and the expansion of the action in these configurations the soft mode action.

\paragraph{Saddle point equation.}
We start by differentiating the action with respect to $B_x$ as
\begin{equation*}
 \left.\frac{\delta S[B]}{\delta B_x}\right|_{B=B^*}=	\frac{2N}{W^2} B^*_x - \frac{\i N}{2} \left(\tr\!_\textrm{o}\rnd{\i B^*_x -  H_0}^{-1} \right)_{xx} \eqd 0\,,
\end{equation*}
which leads to the saddle point equation
\begin{equation} \label{eq:SP_equation_B}
	B^*_x = \frac{\i W^2}{4} \left(\tr\!_\textrm{o}\rnd{\i B^*_x -  H_0}^{-1} \right)_{xx} \,.
\end{equation}
On the right hand side we encounter the resolvent $(\i B-H_0)^{-1}$, which
is a matrix in all but layer indices of the model, evaluated at coinciding sites $x$ and traced over the unit cell. The symmetries of the theory suggest a
site-uniform ansatz for $B^*$ such that $[H_0,B^*]=0$. Together with the
antisymmetry constraint $B = -B^\T$, this implies that a non-trivial stationary
point $B^* \neq 0$ has to possess a minimal  structure in replica space. The
simplest ansatz under these conditions is
\begin{equation}
	B^* = \lambda \tau_2 \,,
\end{equation}
where $\tau_2$ is a Pauli matrix in replica space and $\lambda \in \R$ is the
saddle-point strength. At this point we understand why we had to work with a
replica space of doubled dimension, $2R$: it is required to encode the
symmetries of the microscopic model, $H=-H^\T$, in an inherited symmetry of the integration variables $B$. 

The spatial uniformity of the saddle point ansatz and the translational symmetry
of the Hamiltonian suggest evaluating the saddle point equation in Fourier (momentum) space. Inserting the ansatz above into Eq.~\eqref{eq:SP_equation_B} and
taking a momentum space trace as the integral over the first Brillouin zone 
\begin{equation*}
    \int_q \coloneqq 
    \int_{BZ} \frac{\dd q_1 \dd q_2}{(2\pi)^2}, 
\end{equation*}
with $\int_q 1 = 1$ on both sides of the equation, we obtain 
\begin{equation*}
	\lambda \tau_2 = \frac{W^2}{4} \tr\!_\textrm{o} \int_q \big(\lambda \tau_2 + \i H_0 (q) \big)^{-1} \,.
\end{equation*}

For weak disorder, the integral on the right hand side will be dominated by
low-lying eigenvalues of $H_0(q)$, justifying the evaluation of the cell-trace in
the two-band approximation discussed before (cf.\ Sec.~\ref{sec:haldane-chern}). We thus reduce the Hamiltonian to $H_0 (q) \mapsto H^{(2)} (q) = h_a (q) \sigma_a$, define $h^2 \coloneqq h_a h_a $ and use that $\tr\!_\textrm{o} \,H^{(2)} = 0$ to obtain the self-consistent equation
\begin{equation} \label{eq:SCBA}
	\lambda = \frac{W^2}{2} \int_q \frac{\lambda}{\lambda^2 + h^2} \,.
\end{equation}

Physically, the parameter $\lambda$  plays the role of a self-energy generated
by the disorder, and Eq.~\eqref{eq:SCBA} computes the strength of this
parameter in the self-consistent Born approximation. 
In addition to the trivial solution $\lambda=0$, the equation possesses a non-trivial solution $\lambda \neq 0$, provided the disorder variance exceeds a threshold $W_c(m)$ that depends on the size of the gap $m$. Referring for a detailed discussion to the pioneering
reference~\cite{Fradkin1986} and to Appendix~\ref{app:sp}, we remark that at the
critical point, $m \to 0$, the critical disorder strength $W_c(m)$ vanishes. That is, even for weak disorder there always exists a non-trivial solution,
provided the theory is close enough to the critical point. For $W$ close to $W_c$, this solution is well approximated by
\begin{equation}
\lambda \propto \sqrt{\frac{W}{W_c}-1} \;.
\end{equation}

\paragraph{Symmetries.} 
The symmetry of the action under
$\textrm{O}(2R)$-transformations mentioned above implies the same symmetry for
the stationary phase equation: it is solved by any configuration $B^* \mapsto T
B^* T^{-1}$, where $T^{-1}=T^\T$, as can be checked by direct inspection of the equation. A subgroup
$K$ of these transformations defined by $T \tau_2 T^{-1}=\tau_2$ commutes with
the saddle point and hence does not generate new solutions. For general real
invertible matrices $T\in \textrm{GL}(2R)$, the equation  $T \tau_2
T^{-1}=\tau_2$, defines the real symplectic subgroup, $\textrm{Sp}(2R,\R)$. We
conclude that the intersection $\textrm{O}(2R)\cap \textrm{Sp}(2R,\R)\simeq
\textrm{U}(R)$\footnote{Here, $\textrm{U}(R)$,
defined by its Lie algebra commutation relations, is realized in terms of
\textrm{real} matrices.},
the maximum compact subgroup,
defines the residual symmetry of transformations, leaving the action \emph{and} the saddle point invariant. All other transformations in $\textrm{O}(2R)$ generate \emph{new} saddle points and hence define the Goldstone mode manifold $G/K=\textrm{O}(2R)/\textrm{U}(R)$ of 
a spontaneous symmetry breaking scenario\cite{zirnbauer1996riemannian}. We finally note that a convenient as well as physically motivated representation of the Goldstone mode manifold is
\begin{equation}
    B = \lambda Q \coloneqq 
    \lambda T \tau_2 T^{-1}\;,	\label{eq:B} 
\end{equation}
i.e., the `orbit' of the minimal saddle point generated by the action of symmetry group transformations.

\subsection{Continuum model} \label{sec:continuum} Continuing with the Goldstone
mode program, we now relax the condition of exact commutativity $[T,H_0]=0$ by
allowing for a weak spatial dependence $T\to T_x$ of the symmetry
transformations on the site indices. Anticipating that physically dominant configurations of least action will fluctuate on length scales far exceeding the
unit lattice spacing,  it is expedient to think of the tensor network as
embedded into a continuous two-dimensional plane and replace the site index $x$ by a continuous
two-dimensional variable $x = (x_1,x_2) \in \R^2$. Our Goldstone modes $Q\to
Q(x)=T(x)\tau_2 T(x)^{-1}$ then become continuum fields, and the
non-commutativity $[H_0,Q]\not=0$ will manifest itself in the appearance of
spatial gradients $\partial_{x_\mu} \!Q(x)$ acting on these configurations. Our task now is
to expand the action to leading non-vanishing order in these gradients.

The quadratic term in the action $\sim \tr(B^2)\propto  \tr(T
\tau_2 T^{-1})^2\propto R$ evaluates to a constant (vanishing in
the replica limit) on the saddle point manifold and can be neglected. Defining the (inverse of the) Green's function $G^{-1}=\i \lambda \tau_2 - H_0$ we rewrite the trace-log term in Eq.~\eqref{eq:action_B_trln}  as 
\begin{equation}
 S= -\frac{N}{2} \tr \ln \rnd{1 - G[T^{-1},H_0]T}  + \text{const.}
 \label{eq:S_G}
\end{equation}
In the continuum limit, the commutator $[T^{-1},H_0]$ produces gradient terms of the slowly varying fields $T$. Expanding the action up to second order in commutators/gradients we obtain
\begin{equation}
    S = \frac{N}{2} \rnd{ \tr  \rnd{G [T^{-1},H_0]T} + \frac{1}{2} \tr  \rnd{G [T^{-1},H_0]T}^2 } + \ldots . \label{eq:trln_2nd}
\end{equation}

\paragraph{Gradient expansion.}
To evaluate the expression above, we switch to Wigner's \emph{phase space} representation (for a textbook reference see, e.g., Ref.~\cite{Altland2023}). In the phase space formalism we represent operators $\hat A$ acting in Hilbert space by their \emph{Wigner-Weyl symbols} $A(x,q)$ --- functions of position $x$ and (relative) momentum $q$ in phase space. The transition between operators and symbols proceeds via the Wigner transform 
\begin{equation*}
  \mathcal W[\hat A](x,q) \coloneqq  \int \dd^d y\, \e^{\tfrac{i}{\hbar} q \cdot y}\,   \bigl\langle x - \tfrac{y}{2} \bigr| \, \hat A \, \bigl| x + \tfrac{y}{2} \bigr\rangle.
\end{equation*} 
Where no confusion arises, we will use the notation $A(x,q)\coloneqq \mathcal W[\hat A](x,q) $.

The phase-space representation is beneficial for two reasons. First, we can make use of the fact that, due to translation invariance, the Wigner symbol of $\hat H_0$ (and likewise $\hat G$) reduces to a function of relative momenta $q$ only, while $T$ is diagonal in site-space and hence its symbol is a function of $x$ only. Second, the Wigner transform turns the operator product into the \emph{Wigner-Moyal product} or $\star$-product which explicitly features derivatives in $x$ and $q$ acting on the respective Wigner symbols, i.e.,
\begin{equation}\label{eq:star_def}
    \mathcal W[\hat A \hat B] =
  \mathcal W[\hat A]\star  \mathcal W[\hat B] \coloneqq A(x,q) \exp \rnd{{\frac{i\hbar}{2}\bigl(\larr{\nabla}_x \cdot \rarr{\nabla}_q - \larr{\nabla}_q \cdot \rarr{\nabla}_x\bigr)}} B(x,q). 
\end{equation}
Here, the arrow direction indicates on which symbols the gradients operate. To second order in the gradients (or equivalently in powers of $\hbar$) we obtain
\begin{align*}
  \mathcal W[\hat A \hat B] &= A\,B
    + \frac{\i\hbar}{2}\{A,B\}_{\rm PB}
    - \frac{\hbar^2}{8}\bigl(\partial_x^2 A\,\partial_q^2 B - 2\partial_x\partial_q A\,\partial_q\partial_x B + \cdots\bigr)
    + O(\hbar^3) \,, 
\end{align*}
where $\{A,B\}_{\rm PB} = \partial_x A \, \partial_q B - \partial_q A\, \partial_x B$ is the classical Poisson bracket. Using the straightforward trace relation, $  \Tr \hat A = \int \frac{\dd^d x\,\dd^d q}{(2\pi\hbar)^d}\; \mathcal W[A](x,q)$, applying Eq.~\eqref{eq:star_def} iteratively and expanding to second order in gradients we arrive at 
\begin{align}
	S = -\frac{N}{2} \bigg(
	\underbrace{\i \tr\! \rnd{G\del_\mu H_0\, \Phi_\mu}}_{S_1}
	+ \underbrace{\frac{1}{2} \tr\! \rnd{G \del_\mu \del_\nu H_0 \, \Phi_\mu \Phi_\nu} }_{S_{12}}
	+ \underbrace{\frac{1}{2} \tr\! \rnd{G \del_\mu H_0 \Phi_\mu}^2}_{S_{22}} \bigg) \,,\label{eq:SFJ}
\end{align}
where we have introduced the abbreviation $\Phi_\mu(x) \coloneqq T^{-1}\del_\mu T$ and used the convention that the derivatives $\del_\mu$ are understood as $\del_{x_\mu}$ if acting on $T(x)$ and $\del_{q_\mu}$ if acting on $H_0(q)$. For more details on the derivation we refer to Appendix~\ref{app:Moyal}.

The expression above is still formal inasmuch as it contains momentum integrals
over the $q$-variables hiding in the traces. Carefully separating the momentum-
and position-dependent variables (as outlined in detail in Appendix~\ref{app:GradientExpansion}) leads to our main result, the continuum action (cf.\ Eq.~\eqref{eq:Scont}) known as the
Pruisken nonlinear $\sigma$-model \cite{LevineQHE:1}
\begin{equation} 
 \begin{split}
 	S[Q]
 	&= \frac{N}{2} \,\bigg( \underbrace{g \int_x \tr(\del_\mu Q \del_\mu Q)}_{\Sgrad [Q]} +  \underbrace{\frac{\vartheta}{16 \pi} \int_x \ltop}_{ \Stop [Q]} \bigg) \,,
 \end{split}
\end{equation}
with couplings $g$ and $\vartheta$ given by the momentum integrals stated in
Eq.~\eqref{eq:bare-couplings}. 

{
\paragraph{Source terms.} In a final construction step, we recall the presence
of the source matrix $J$ in the original action, which we had temporarily
suppressed for notational brevity. Restoring it in the action
Eq.~\eqref{eq:S_G}, it features as an additional term in the logarithm, 
\begin{align}
    S[Q,J]=-\frac{N}{2} \tr \ln \rnd{1 -
G[T^{-1},H_0]T + G T^{-1} J T}+\text{const.}
\end{align}  
To make the discussion  simple, let us restrict the expansion to order
$\mathcal{O}(J^2)$, as required for the computation of the first and second moments
of our correlation functions Eq.~\eqref{eq:CorrelationFunctionSource}. In the
expansion in $J$, we focus on terms of dominant order, $\sim J Q$, neglecting
spatial gradients present in the commutator $[T^{-1},H_0]$. With this
simplification, we obtain 
\begin{align}
    S[Q,J]\approx S[Q]- \frac{N}{2}\tr(TGT^{-1}J)+\frac{N}{4}\tr(TGT^{-1}J)^2 + \ldots \,.
\end{align} 
Expanding to second order in sources $J$ and keeping the leading \emph{local} mean-field contribution $T G T^{-1} = \allowbreak \cst. \times Q$, the linear term vanishes for off-diagonal sources with $x\!\neq \!y$, while away from this approximation it remains exponentially suppressed at large separations. The quadratic term yields a bilocal contribution
\begin{align}
    S^{(2)}[Q,j]\coloneqq \mathrm{const.}\times \sum_{m,n=1}^{R} j^m j^n \tr\!_\tau \Big(Q(x)^{mn} Q(y)^{nm} \big) \,,
\end{align}
where $Q^{mn}$ are $(R\times R)$ replica blocks and $\tr\!_\tau$ is taken over the replica-doubling space.
With this result, the computation of the correlation function moments according
to Eq.~\eqref{eq:CorrelationFunctionSource} yields
\begin{align}
   &\mathbb{E} \sqd{\mathcal{C}(\theta_{x},\theta_{y})} = 0 \,, \qquad 
   \\
    &\mathbb{E} \sqd{\mathcal{C}(\theta_{x},\theta_{y})^2}  \propto 
    \lim_{R\to 0} \av{ \tr\!_\tau \rnd{Q^{12}(x) \,Q^{21}(y) + (1\leftrightarrow 2) } }_{Q} \,, 
\end{align} 
where the superscripts refer to the $R$-replica labels and the trace is over the $\tau$-doubling.
The first of these equalities states the vanishing of first moments of non-local Grassmann
correlation functions (at mean-field level) due to random sign fluctuations weighting realization-specific matrix elements
of the correlation matrix $H^{-1}$. 
while the second computes the \emph{variance} of these
correlation functions in terms of a functional expectation value of the
$Q$-field\footnote{More explicitly, for a fixed realization the nonlocal matrix element $H^{-1}_{xy}$ can be viewed as a sum of many disorder-dependent contributions with alternating signs, whose average cancels to leading order; meanwhile, products of two such amplitudes retain positive diagonal pairings and therefore survive disorder averaging.}.  
}

\paragraph{Generator expansion.} \label{sec:generators_free} For values of the
coupling $gN\gtrsim \mathcal{O}(1)$, fluctuations of the $Q$-fields are limited
to small deviations away from the saddle point $Q=\tau_2$.  With
$T=\exp(W/2)$ such fluctuations may be described by expansion of the action to
leading order in generator matrices $W= X_a \tau_a = X_1 \tau_1+ X_3\tau_3$, chosen to anticommute with $\tau_2$, and antisymmetric $X_a = -X_a^\T$.  
Substituting this representation into Eq.~\eqref{eq:Scont}, the straightforward expansion to second order in $X$ yields Eq.~\eqref{eq:S_pruisken_gen}. {For off-diagonal replica blocks $m \neq n$ ($m,n \in 1,\dots,R$), we have $Q^{mn} (x)  \sim \Ocal(W^{mn})$. Thus, to the same leading order of the expansion, the correlation function reduces to 
\begin{align}
    \label{eq:CorrelationFunctionX}
    \mathbb E \left[ \mathcal{C}(\theta_{x},\theta_{y})^2 \right] &\propto
     \lim_{R\to 0} \av{ \rnd{ X_1^{12} (x)\, X_1^{21} (y) + X_3^{12}(x)\, X_3^{21} (y)} + (1 \leftrightarrow 2)}_{X}.
     \end{align}  
Doing the Gaussian integral, we obtain the result Eq.~\eqref{eq:FourPointPlane}.
Note that these remain unaffected
by the topological term (at this perturbative level), reflecting its nature as a total derivative.}


\section{Quartic terms}
\label{sec:quartic}

In the previous section, we have derived an effective continuum description for an ensemble of random matchgate tensor networks. A natural question to ask now is how this continuum theory is modified by going beyond matchgates --- equivalently, in the fermionic picture, beyond a free (Gaussian) theory. In a prior work \cite{WilleQuarticTN:2025} we have introduced a minimal non-Gaussian extension of an identical tensor network in terms of a site-local quartic term which adds to the action as  $S[\theta] =S_{\tx{free}}[\theta] + S_{\tx{int}}[\theta]$, with the interacting action given by
\begin{equation}
    \label{eq:QuarticTheta}
    S_{\tx{int}} = -b \sum_x \theta_{1,x} \theta_{2,x} \theta_{3,x} \theta_{4,x}
   \,.
\end{equation}
Here, $b$ is a real-valued coupling constant.

\paragraph{Two-band approximation.}
As in the free theory, we start by subjecting this action to a  two-band approximation. However, as the quartic terms couple the low and the high lying bands, the naive projection employed so far is no longer sufficient. Instead, we consider the quartic term in the full eigenbasis of the clean band Hamiltonian $\theta^\alpha_{q} = V^{\alpha\T}(q) \eta_q$ and obtain four types of terms: purely low-band couplings, purely high-band couplings and mixed terms. First, we integrate out the higher bands. The pure high-band couplings turn into an irrelevant constant, while the mixed terms generate a `renormalization' of the free Hamiltonian as detailed in Appendix B of Ref.~\cite{WilleQuarticTN:2025}. For weak enough $b$, this leads to a shift of the phase boundary, but does not change the free Hamiltonian qualitatively. 

Turning to the purely low-band quartic terms,  we observe that these vanish if we neglect the momentum dependency of the band-vectors (i.e., setting $V(q) \to V(0)$), since, in this approximation, we deal with strictly local terms containing  two (or more) copies of the same
Grassmann mode. However, if we keep the momentum dependency, the momentum space products $V^{\T}(q) \eta_q$ become real-space convolutions $\sum_y K(x-y) \eta_y$  with short-ranged kernels, and as a result the strictly local quartic $\theta$-term above becomes a \emph{quasi-local} quartic term of low-lying modes.
Neglecting the microscopic details, we approximate this term by its generically dominant contribution: the nearest-neighbor pair interaction $\sum_r \sum_{\langle x,y \rangle} (\eta^r_1 \eta^r_2)_{x} (\eta^r_1 \eta^r_2)_{y}$. We here include the sum over replica indices to emphasize that this interaction term breaks the replica rotation symmetry $\eta^r \to T^{rs}\eta^s$ fundamental to our theory; whatever emerges from these terms after the integration over the $\eta$-fields will be a strong intruder into the structure of the theory. This being so, we will put the focus on identifying the structure of these modifications, but not on their detailed quantitative form. 

\paragraph{Decoupling.}
In order to prepare the quartic term above for a Hubbard-Stratonovich decoupling akin to the procedure in Sec.~\ref{sec:replica}, we introduce the lattice bilinears (`densities') $\rho^r_{c,x} = \frac{1}{2} \eta^{r\T}_{c,x} (\i\sigma_2) \eta^r_{c,x}$, where $c = 1,\dots,N$ denotes the layer index. In this representation, the quartic action assumes
the form 
\begin{equation}
	S_{\tx{int}} \simeq -\frac \gamma 2 
    \sum_{r,c}\sum_{\langle x,y \rangle} \rho^r_{c,x} \rho^r_{c,y}= -\frac{\gamma}{2}\rho^\T M \rho  \,, \quad \tx{with} \quad \gamma \propto b \,,
\end{equation}
where the second equality introduces a vector-matrix notation, 
\begin{equation}\rho^\T M
\rho=\sum_{r,c,x,y} \rho^r_{c,x} M_{xy} \rho^r_{c,y}, 
\end{equation}
and $M_{xy}=1$ if $x$ and $y$ are
nearest neighbors and zero otherwise (in particular, $M_{xx} = 0$ for all sites). Introducing a local field of real scalar
variables $Y=\{Y^r_{c,x}\}$ we decouple this bilinear in terms of a Hubbard-Stratonovich transformation
\begin{equation}
	\e^{-S_{\tx{int}}} \propto \int \dd Y \exp \left(-\frac{1}{2\gamma} Y^\T M^{-1} Y + Y^\T \rho \right)\,
\end{equation}
where $M^{-1}$ is the inverse of $M$, which we will not need to evaluate. Partially unraveling the notation as 
\begin{equation}
Y^\T \rho = \sum_{r} Y^{r\T}\rho^r=\frac{\i}{2} \eta^{r\T} Y^r \sigma_2 \eta^r,
\end{equation}
we obtain the generalized fermionic action $S[\eta,Y] = -\frac{\i}{2} \eta^{r\T} (H + Y^r \sigma_2) \eta^r$. From this point onward, the construction of the theory proceeds as in the non-interacting case, the only difference being the presence of the $Y$-field. Assuming the interaction strength, $b$, to be sufficiently weak to not alter the mean field solutions, the ensemble averaged system is described by 
\begin{align}
    S [\eta,Y,Q]&= -\frac{\i}{2}
\eta^{\T} (- \i \lambda Q + H + Y \sigma_2) 
\eta +\frac{1}{2\gamma} Y^\T M^{-1} Y \,,
\\
   \stackrel{\int \dd \eta}{\longrightarrow} \;\; 
   S[Y,Q] &= -\frac{1}{2} \tr \ln \rnd{\i \lambda Q - H - Y\sigma_2}+\frac{1}{2\gamma} Y^\T M^{-1}Y,
\end{align}
where in the last step we integrated over the fermions, and $H$ now represents the clean part of the correlation matrix.

\begin{figure}
	\centering
	\includegraphics[width=0.5\linewidth]{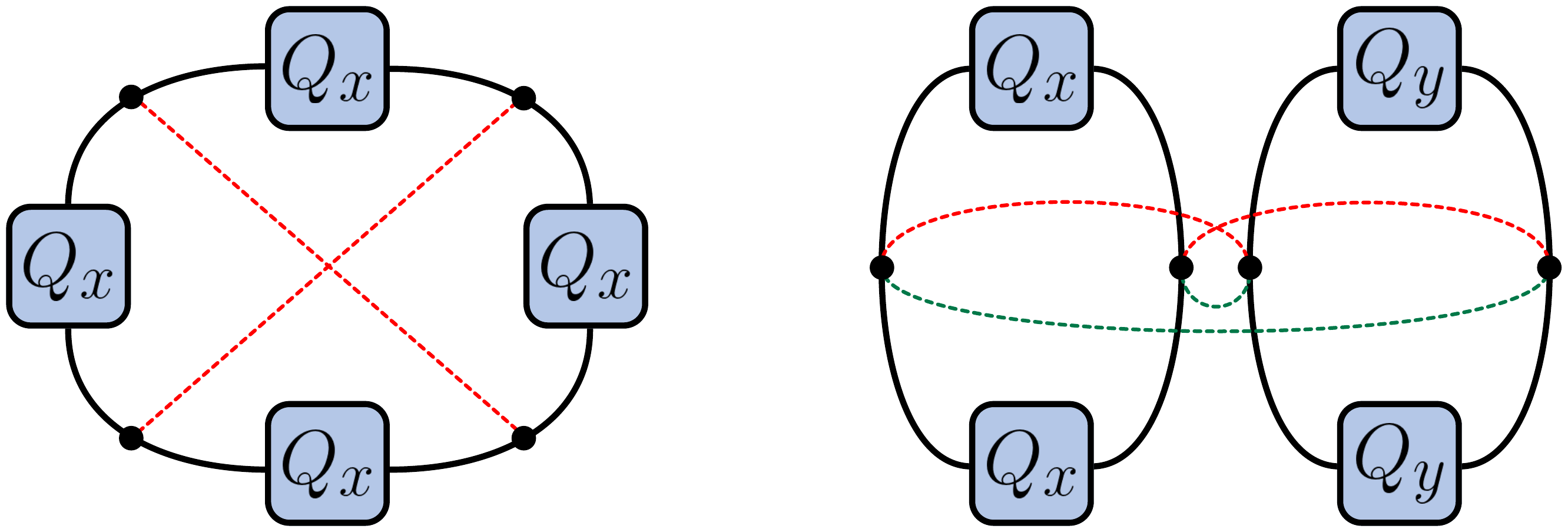}
	\caption{Diagrams corresponding to the two Wick averages in
	Eq.~\eqref{eq:exp_two_avs}, with layer indices suppressed for brevity. Black
	lines indicate index contractions in replica space, black dots are the
	fields $Y^r_x$ with the corresponding replica index. Dotted lines illustrate
	contractions caused by taking the average $\av{\dots}_Y$ (those permitted 
    by
	the replica structure). Left: The average of $\tr(QY)^4$ vanishes due to
	$\av{Y_x Y_x}=0$. 
    Right: The average of $(\tr(QY)^2)^2$ allows for two
	(equal) 
    non-vanishing contractions. }
	\label{fig:avs}
\end{figure}

\paragraph{Expansion in the weak quartic limit.}
From here we  proceed by expansion of the tr ln in the parameter $\gamma/\lambda$.
Using the stationary phase identity Eq.~\eqref{eq:SP_equation_B} evaluated on the Goldstone manifold for a replacement 
\begin{align}
    Q_x \approx \i\lambda \left( \i\lambda Q  -H \right)^{-1}_{xx},
\end{align} 
and executing the free running trace over even powers of the two-band matrix $\sigma_2$, we obtain the effective action 
\begin{align}
    \e^{-S_\mathrm{int}[Q] }\coloneqq \left\langle \exp 
      \left( - \frac{1}{2} \sum_{m=1}^{\infty} \frac{(-1)^m}{m \lambda^{2m} }  \tr(Q Y)^{2m} \right)  \right\rangle_Y,
\end{align} 
where $\langle \dots \rangle_Y$ denotes averaging over the Gaussian $Y$-weight,
\begin{align}
    \langle Y^r_{c,x} Y^s_{d,y} \rangle_Y = \gamma \delta^{rs}_{cd}M_{xy},
\end{align}
and higher moments are computed via Wick's theorem. Our $Q$-matrices are replica off-diagonal, implying that the average of lowest order $\langle \tr(QYQY) \rangle_Y\propto \sum_r Q^{rr}Q^{rr}$ vanishes. To leading non-trivial order, we thus obtain
\begin{equation} \label{eq:exp_two_avs}
    \e^{-S_\mathrm{int}[Q] }\approx \exp \left(  \frac{1}{4 \lambda^4}\left( 
        \frac{1}{2}  \langle (\tr(QY)^2)^2 \rangle_Y - \langle \tr((QY)^4) \rangle_Y \right)   \right) \,.
\end{equation} 
Looking at these two contributions separately (cf.\ diagrams in Fig.~\ref{fig:avs} for the replica contractions), we find 
\begin{align}
   \langle \tr((QY)^4) \rangle_Y & = \left\langle \tr(Q_x^{rs} Y^{s}_{c,x} Q^{st}_x Y^{t}_{c,x} Q^{tu}_x Y^{u}_{c,x} Q^{ur}_x Y^{r}_{c,x}) 
   \right\rangle_Y=  0, \quad \tx{and} 
   \\
    \langle (\tr(QY)^2)^2 \rangle_Y &= 
    \left\langle (Q_x^{rs} Y^{s}_{c,x} Q_x^{sr}Y^{r}_{c,x})(Q_y^{rs} Y^{s}_{d,y} Q_y^{sr}Y^{r}_{d,y}) \right\rangle_Y= 
    2 \gamma^2 N \sum_{\langle x,y \rangle} (Q_x^{rs} Q_x^{sr})(Q_y^{rs} Q_y^{sr}), 
\end{align} 
where the vanishing of the first term follows from a conflict between the site-diagonal structure of the $Q$-field and the nearest-neighbor structure of the $Y$-correlation --- namely, recall that $\avs{Y_x Y_x} \propto M_{xx} = 0$. The factor of $N$ appears due to a single free running summation over $c$-layer indices. Finally, using that $Q_x \approx Q_y$ in the slow field limit, the last term can be simplified to $\approx 8N \gamma^2 \sum_x (Q_x^{rs}Q_x^{sr})^2$. Substituting this expression back into our exponent, we obtain the (local) interaction contribution to the action
\begin{align}	
	\label{eq:InteractionReplicaProjectors}
    S_\mathrm{int}[Q] =  \gamma^{(4)} \sum_x \tr_{\tx r} \,(Q_x P^r Q_x P^s Q_x P^r Q_x P^s)\,, \with \gamma^{(4)} \propto \frac{ N \gamma^2}{\lambda^4} \,.
\end{align}
Here, we have introduced the replica projectors $P^r_{st} = \delta_{rs} \delta_{rt}$,
and a summation over replica indices is implicit. 

\paragraph{Generator expansion.} Upon promoting site-local matrices $Q_x$ to
continuum variables $Q(x)$, this term modifies the continuum action in
Eq.~\eqref{eq:Scont} by a local quartic term no longer invariant under
continuous transformations $Q\to T Q T^{-1}$ in replica space. To make its
impact on the low-energy properties of the theory explicit, we may parameterize
the $Q$ fields in terms of the fluctuation generators $X_{1,3}$ previously used
in Section~\ref{sec:continuum}. A straightforward expansion then 
shows that the lowest
non-vanishing contribution in generators appears at \textit{quadratic} order in
$X$:
\begin{equation}
    \label{eq:InteractionGeneratorExpansion}
	S_{\text{int}}[W] \simeq \gamma^{(4)} \int_x \tr \rnd{X_1^2 + X_3^2} + \Ocal(X^4) \,.
\end{equation}
Being quadratic in generators, and not involving gradients this is a highly
relevant perturbation to the action, turning previously long ranged correlations
into ones decaying exponentially in the ratio $\sim \gamma^{(4)}/g$. 

\paragraph{Replica permutation symmetry.} However, this damping effect on long
range fluctuations does not directly imply that interactions confine $Q$ to stay close to
the chosen saddle point $\tau_2$. The reason is that our quartic term still
respects the discrete replica permutation symmetry $Q \to P Q P^{-1}$, where $P$ is a
permutation matrix acting in replica space. (Technically, such permutations
change the indices $r,s$ of the projector matrices in
Eq.~\eqref{eq:InteractionReplicaProjectors}	into permuted indices, which remains
inconsequential upon summation.) This residual symmetry turns our continuum
field theory into a discrete symmetry `statistical mechanics' model. The
permutation degrees of freedom of the latter play an important role in the
analysis of, e.g., entanglement properties, as we will discuss in future work.

We finally note that the inclusion of  nonlinearities into a disordered tensor
network, need not necessarily demote the continuum theory to a discrete one.
What all nonlinearities have in common is that they violate continuous replica
rotation symmetry (as a consequence of the non-invariance of terms $ \sim
\theta^n$, $n>2$.) However, one can easily construct nonlinearities invariant
under other symmetry groups. As an example, consider the bilinear bond
correlation operators $X_{x,y}\coloneqq \sum_{c=1}^N \theta_{x,c}\theta_{y,c}$
squared to $\sum_{\langle x,y \rangle} X^2_{x,y}$. This realizes a quartic term
with continuous $\textrm{O}(N)$ symmetry under rotations of the layer index $c$,
coexisting with replica permutation symmetry. The example shows that there is
rich playground for the engineering of discrete and continuous symmetry
tensor networks subject to nonlinear correlations and randomness.


\section{Continuum theory on the hyperbolic plane}
\label{sec:hyperbolic}

{
It is conceptually straightforward to extend the continuum description to
non-Euclidean domains. Motivated by their potential relevance to holographic
constructions, tensor networks defined on tessellations of the hyperbolic plane
have attracted significant attention in recent years~\cite{Pastawski_2015,Evenbly_2017,HolographicReview}. Referring to the
forthcoming Ref.~\cite{hyperbolicAL2026} for details, feeding such geometries into the derivation of
our continuum theory by gradient expansion yields a generalization of the action
in Eq.~\eqref{eq:Scont} whose structure is dictated by geometric constraints: the
previously planar area element is replaced by the canonical area element of the
domain space, $\int_x = \int \dd^2 x \mapsto \int \dd^2 x\sqrt{g(x)}$, where $g(x)$ is the determinant
of the metric tensor, depending on two coordinates $x=(x^1,x^2)$; the gradient
term, $-\frac{Ng}{2}\int_x \tr(Q \Delta Q)$\footnote{The Laplacian appears straightforwardly by partial integration in Eq.~\eqref{eq:Scont}.}, now contains the Laplace-Beltrami
operator
\begin{align}
    \label{eq:LaplaceBeltrami}
    \Delta = \frac{1}{\sqrt{g}}\,\del_\mu
    \bigl(\sqrt{g}\, g^{\mu\nu} \del_\nu\bigr).
\end{align}
Finally, the topological term remains form-invariant, owing to its metric
independence (formally, the change of the area element cancels against the
determinant picked up by the antisymmetric derivative combination). 
Focusing on
the large-$g$ thermal metal regime, we are thus led to consider
\begin{align}
    \label{eq:ActionHyperbolic}
    S[Q]= -\frac{Ng}{2} \int \dd^2 x \sqrt{g}\,\tr(Q \Delta Q).
\end{align}

There are various convenient coordinate systems on the hyperbolic plane, each
having its own advantages. In view of the fact that tensor networks are often
studied numerically on domains of finite radius $R$ relative to some center
point, we here choose to work in polar coordinates $(r,\phi)$, where $r$ is the
geodesic distance from the center and $\phi$ the angular coordinate. The metric
in these coordinates is given by
\[
\dd s^2 = \dd r^2 + \sinh^2(r)\, \dd \phi^2,
\]
where the curvature radius has been set to unity for simplicity. In these
coordinates, it holds $\sqrt{g}=\sinh(r)$, demonstrating the reduction to the planar area
element $r\,\dd r\,\dd \phi$ for $r\ll1$, and exponential area increase
$\sim \e^{r}\,\dd r\,\dd\phi$ for large radii.

It is interesting to study what this change of geometry implies for the behavior
of correlation functions. This can be seen as a continuum version of the explorations of the decay of correlations on the boundary depending on the curvature in the bulk
\cite{Jahn}. To this end, consider the linearization 
\begin{align}
    S[X] = Ng \int \dd^2 x \sqrt{g}\,\tr(X_a \Delta X_a)
\end{align}
of the $Q$-field theory in terms of generators $X_a = X_1, X_3$ (recall that $Q = \tau_2 \exp(X_a \tau_a)$).
The two-point correlation function of the generators
Eq.~\eqref{eq:CorrelationFunctionX} is then given by the 
Green's function of the
Laplace--Beltrami operator, $\langle x|\Delta^{-1}|y\rangle$. These matrix
elements are known~\cite{Helgason1984}, and in the following we walk through a
number of resulting predictions for tensor network correlation functions
$F(x,y)\propto \frac{1}{gN}\langle x|\Delta^{-1}|y\rangle$, cf.\ Eq.~\eqref{eq:FourPointPlane}.

For an infinite hyperbolic plane, the correlator reads (up to an overall sign
convention)
\[
\avs{ x| \Delta^{-1} |y }
= \frac{1}{2\pi}\ln\tanh \bigl(d(x,y)/2\bigr)\,,
\]
and it depends only on the geodesic distance
$d(x,y)$ between the observation points. For example, setting $y=0$ and
$x=(r,\phi)$ one obtains
\begin{align}
    \label{eq:CorrelationFunctionHyperbolic}
    F(x,0)\propto \frac{1}{gN}\ln\tanh \,\left(\frac{r}{2}\right).
\end{align}
This expression interpolates between the planar logarithmic behavior
$\sim \ln(r)$ for $r\ll1$ and an exponentially small value
$\sim \e^{-r}$ for $r\gg1$. The latter reflects the exponential volume growth of
the hyperbolic plane, which suppresses return probabilities at large distances.

Another interesting configuration is $y=(R,0)$ and $x=(R,\phi)$, corresponding
to correlations between points of angular separation $\phi$ on a circle of
radius $R\gg1$, chosen larger than the curvature radius to make non-Euclidean
effects manifest. From the hyperbolic law of cosines one finds, for $R\gg1$,
\[
d(x,y)\approx 2R + 2\ln|\sin(\phi/2)|,
\]
which leads to a correlation function
\begin{align}
    \label{eq:CorrelationFunctionRing}
    F \rnd{ (R,\phi),(R,0) } \propto \frac{1}{gN}\,
    \frac{\e^{-2R}}{\sin^2(\phi/2)},
\end{align}
exponentially small in the ring radius.

An interesting feature of these correlations is their extreme sensitivity to
boundary conditions. Truncating the plane at radius $R$ (as required, for
example, in numerical simulations) imposes Neumann boundary conditions on the
Laplace inversion,
\[
\partial_r \Delta^{-1}(x,y)\big|_{r=R}=0.
\]
The boundary-restricted correlation function then loses its dependence on the
bulk geodesic distance and becomes logarithmic in the angular separation,
\begin{align}
    \label{eq:CorrelationFunctionRingNeumann}
    F((R,\phi),(R,0))
    \propto \frac{1}{gN}\,\ln \,\bigl(\sin(\phi/2)\bigr) + C(R),
\end{align}
where $C(R)$ is an additive constant depending on the normalization of the
Neumann Green's function. Compared to the infinite-plane result, this represents a
parametrically strong enhancement of boundary correlations enforced by the
truncation of the geometry.
These examples illustrate typical large-distance correlations in matchgate
tensor networks arising purely from continuum geometry. It would be interesting
to compare these predictions to numerical simulations on lattice realizations
of hyperbolic domains.}


\section{Conclusion}
\label{sec:conclusion}

In this work, we have introduced a way of capturing continuous models making use of random tensor networks, invoking a notion of typicality.
Concretely, 
this work establishes a continuum, long-wavelength description for
\emph{ensembles} of two-dimensional matchgate tensor networks. Once tensor
parameters fluctuate spatially, the relevant description shifts from
microscopic tensor data to macroscopic, universal structures governing behavior
on large length scales. In close analogy with disordered fermionic systems,
randomness acts as a mechanism of universality: many distinct microscopic
realizations share the same large-scale phenomenology, controlled primarily by
symmetry and topology. From a quantum-information perspective, this provides a
principled route to classifying Gaussian tensor-network ensembles with
fluctuating parameters in terms of a small set of continuum data, most notably a
transport parameter $g$ and a topological angle $\vartheta$.
{Within this framework, moments of fermionic two-point functions become field-theoretic observables that directly probe the long-distance physics.}

A central conceptual ingredient is the reorganization of the disorder average
via a Hubbard-Stratonovich transformation. This condensed-matter technique,
still uncommon in tensor-network analyses, replaces an average over fluctuating
tensor parameters by an equivalent description in terms of a collective field in
replica space. In doing so, it renders the long-distance problem amenable to
standard field-theoretic methods and leads to an effective theory for slowly
varying continuum degrees of freedom $x\mapsto Q(x)$. {In particular, disorder-averaged correlators and their fluctuations are mapped to $Q$-field correlation functions.}

The continuum description is quantitatively controlled in a large-$N$ limit,
where $N$ counts the number of coupled layers, and serves as an organizing
framework for admissible universal scenarios at smaller $N$. Our analysis
focuses on a minimal square-lattice ensemble with a single fluctuating parameter
$a$. Via a duality of the type discussed in Ref.~\cite{WilleQuarticTN:2025}, the
associated fermionic kernel $H=-\i(C+A)$ maps to a Haldane--Chern
insulator band structure. This makes the symmetry and topological content
explicit and enables a direct translation between tensor-network data and the
continuum language of class~D disordered systems. {Concretely, the resulting sigma-model correlators reproduce the characteristic long-distance behavior governed by class~D soft modes in the thermal metal regime.}

Furthermore, we show how the framework extends beyond the strictly Gaussian setting in a
controlled weak-deformation regime. Small non-Gaussian perturbations generate a
local, symmetry-reducing term that is quadratic in the fluctuation generators
and therefore strongly relevant. This term acts as a mass for the Goldstone
modes, cutting off long-ranged correlations while preserving replica permutation
symmetry. 

Taken together, the approach functions as a two-way bridge between
discrete tensor network ensembles and continuum field theories: it imports
methods from disordered condensed-matter physics into the tensor-network
setting, while embedding random tensor networks into a continuum classification
by transport and topology. At a broader level, this work brings together 
quantum-information inspired tensor network approaches and quantum field theory methods from condensed 
matter physics.

The continuum formulation developed here suggests several extensions. First,
one may generalize the underlying lattice to non-translationally invariant or
curved tessellations. In the continuum framework this leads naturally to field
theories coupled to background metrics, providing access to long-range
correlation functions in, for example, hyperbolic geometries. 
{As an illustration, we evaluated sigma-model correlators on a hyperbolic disk, where negative curvature reshapes the long-wavelength structure and leads to boundary-dominated correlation profiles.}
{A more detailed discussion of geometric effects on long-distance behavior, including implications for localization, will be provided elsewhere~\cite{hyperbolicAL2026}.}
Second, the
different symmetry-breaking regimes discussed above have direct consequences
for entanglement properties, ranging from area-law behavior in symmetry-unbroken
phases to logarithmic and volume-law entanglement in phases with broken
continuous and discrete symmetries. In this way, the ensemble theory provides a
controlled notion of typicality across backgrounds with distinct entanglement
signatures. 
Finally, the framework can be adapted to problems motivated by quantum
device physics, including the description of discrete-time circuit dynamics
using matchgate tensors with non-Hermitian correlation matrices, or the modeling
of isolated device imperfections such as single-qubit errors via tailored
non-Gaussian disorder distributions.

\section*{Acknowledgements}

We thank Alexander Jahn for discussions. M.U. thanks Mateo Moreno
Gonsalez and
Shozab Qasim for discussions. 
This work has been supported by the Deutsche Forschungsgemeinschaft (DFG, 
German Research Foundation) under
Germany’s Excellence Strategy Cluster 
of Excellence Matter and Light for Quantum Computing (ML4Q) EXC 2004/1 (Project Grant No.\ 390534769), and within the CRC network CRC 183 (Project Grant No.\ 277101999) as part of subproject B04, as well as by BMFTR
(MuniQC-Atoms, FermiQP), the Quantum Flagship (PasQuans2 and Millenion), Berlin Quantum, and the ERC (DebuQC).
Generative AI tools (ChatGPT-5 and higher) were used for text polishing such as grammar and spelling checks.



\newpage
\begin{appendix}
\numberwithin{equation}{section}

\section{Background: Disordered class D superconductor}
\label{sec:phases}
In Secs.~\ref{sec:SummaryOfResults} and \ref{sec:GFTN}, we emphasized the proximity of our tensor-network ensemble to disordered fermionic systems in symmetry
class~D. For the convenience of readers not familiar with the phenomenology of
this symmetry class, we briefly review the main physical ingredients that enter
the continuum description used in this work (cf.\ the nonlinear sigma model
introduced above in Eq.~\eqref{eq:Scont}). For more detail we refer to the literature such as Refs.~\cite{Bocquet2000a,Efetbook,MirlinReview} as well as others cited in the main text.

Being a superconductor, a class~D system admits a \emph{Bogoliubov-de~Gennes} (BdG) representation in terms of complex fermion operators, cf.\ Eq.~\eqref{eq:Bogolubov-deGennes}. Using only the minimal structural constraints on its blocks (Hermiticity of the quasiparticle Hamiltonian $h=h^\dagger$ and
antisymmetry of the pairing $\Delta=-\Delta^\T$) the same BdG Hamiltonian can be
rewritten in a real Majorana basis as a purely imaginary antisymmetric matrix as in Eq.~\eqref{eq:BdGVsMajorana}. In a path-integral formulation, the Majorana
operators become Grassmann fields, and the theory reduces to a bilinear form of
exactly the same structure as the one governing matchgate tensor networks.

Generic crystalline class~D superconductors possess a spectral gap around zero
energy (the BCS gap) and are thus thermal insulators: temperature gradients do
not induce bulk heat conduction. In two spatial dimensions these insulating
states can nevertheless be topological, with integer-valued band invariants
(Chern numbers). When the invariant vanishes, the phase is trivial; otherwise
it realizes a \emph{thermal quantum Hall insulator}. As in the conventional
(class~A) quantum Hall effect, non-trivial topology implies chiral boundary
modes --- here chiral Majorana modes --- responsible for edge heat transport.

Phase transitions between distinct thermal quantum Hall phases necessarily
involve bulk gap closings. In the vicinity of such transition points (both in
energy and in crystal momentum) the band structure can be linearized, yielding a
two-dimensional Dirac description; see Section~\ref{sec:haldane-chern} for the
corresponding construction in our setting. These Dirac theories provide useful
local proxies for the full band structure near criticality.

Disorder (of strength $W$) breaks translational invariance and compromises sharp
band gaps through the emergence of impurity states. For weak disorder, and with
the clean system tuned close to a gap-closing transition, the effective
description is that of Dirac fermions subject to a random mass term (the only
Dirac perturbation compatible with class~D symmetry). Perturbative analyses find
that weak disorder is an irrelevant perturbation in this regime~\cite{Bocquet2000a,Wang2021}: at large length
scales the system appears effectively clean, so the clean criticality survives
into the weakly disordered regime (as reflected by the extension of clean phase
boundaries into the $(W,a)$ plane in Fig.~\ref{fig:phase}).

Remarkably, for disorder exceeding a nonuniversal critical strength, class~D
systems can undergo a transition from insulating behavior to a conducting phase,
the \emph{thermal metal}. The thermal metal is characterized by power-law
(``diffusive'') correlations and a finite density of states at zero energy. In
field-theoretical language, this transition can be phrased as spontaneous
symmetry breaking between disorder-averaged advanced and retarded Green's
functions
\[
G^\pm(\eps)\coloneqq(\eps\pm \i\delta-H)^{-1},
\]
evaluated at the symmetry point $\eps=0$. Here, $\delta$ is an infinitesimal regulator. A convenient diagnostic is the
density of states at zero energy
\[
\rho \coloneqq \frac{1}{2\pi i}\,\big \langle \tr(G^- - G^+)\big \rangle,
\]
which vanishes in insulating phases and becomes non-zero in the thermal metal~\cite{MirlinReview}.

While the existence of the thermal metal can be motivated within the Dirac
picture, its long-distance properties are more efficiently captured by the
nonlinear sigma model framework employed above. In particular, for systems with
large conductance (as realized by our multilayer construction), the coupled flow
of the stiffness and the topological angle in Eq.~\eqref{eq:Scont} provides a
natural language for the competition between localization, quantum Hall
criticality, and thermal metallicity. The characteristic tendency toward a
metallic phase is understood as class~D antilocalization: Goldstone-mode
fluctuations associated with replica symmetry breaking enhance, rather than
suppress, transport~\cite{Bocquet2000a,MirlinReview}.

Finally, as in the closely related field theory of the conventional quantum Hall effect (see Ref.~\cite{MirlinReview} for a review), the detailed critical physics of the thermal quantum Hall transition
itself lies at strong coupling and remains beyond quantitative analytic control.
What can be extracted reliably is the qualitative structure of the flows and the
existence of a finite critical conductance $g^\ast$ at $\vartheta=(2n+1)\pi$, in
agreement with numerical studies~\cite{ChalkerMerz,Wang2021} which find a critical conductance of order
unity.

\section{Fourier transform of a real Grassmann bilinear form}
\label{app:FT}
In the derivation of the momentum space representation of the bilinear form $\theta^\T H_0 \theta$ with $H_0$ defined in Eq.~\eqref{eq:HC}, we use the following conventions and definitions.
We take the Fourier transform with respect to the lattice structure $C$, meaning, we introduce 
\begin{equation*}
\theta_{q,i}=\frac{1}{L} \sum_x \e^{-\i q \cdot x} \theta_{x,i}\;, \qquad  \text{and} \qquad 
H_{ij}(q,p)=\frac{1}{L^2} \sum_{x,y} H_{x i,yj} \,\e^{-\i q \cdot x + \i p \cdot y}\;,
\end{equation*}
where $L^2$ is the number of cells in the lattice, $x,y$ are vectors of unit-cell positions and $q=(q_1,q_2)^\T$, $p = (p_1,p_2)^\T$ are their momentum conjugates. Since $H_{ij}(q,p)=H_{ij}(q)\delta_{q,p}$ we obtain the compact form $\sum_{q} \theta^\T_{-q} H(q) \theta_{q}$, where $H(q)$ is now a $4 \times 4$-matrix known in condensed matter theory as a band Hamiltonian.

Note that this bilinear form couples opposite momenta and we can evaluate the integrals as 
\begin{equation*}
\prod_{q>0} \left(\int \dd \theta_q \dd \theta_{-q} \exp \theta_{-q}^\T H_q \theta_q\right) =\prod_{q>0} \det H_q \;.
\end{equation*}
Using $-H^\T(q)=H(-q)$ and assuming we have an even number of cells in the lattice we find
\begin{equation*}
\int \left(\prod_q  \dd \theta_q   \right) \exp \left( \sum_q \frac 1 2\theta_{-q}^\T H_q \theta_q \right) = \pm \exp \left( \frac 1 2 \sum_q \tr \ln H_q \right) \;,
\end{equation*}
where the global sign factor is irrelevant for any physical observables.

\section{Stationary point analysis}
\label{app:sp}
The saddle point strength $\lambda$ is given by 
\begin{equation} \label{eq:lambda_SCBA}
	\lambda = \frac{W^2}{2} \int_q \frac{\lambda}{\lambda^2 + h^2} \,,
\end{equation}
which is known as the \textit{self-consistent Born approximation} (SCBA)\cite{Altland2023}. Its non-trivial solutions $\lambda \neq 0$ are determined by the self-consistent equation
\begin{equation} \label{eq:integral_eq_sp}
	\frac{2}{W^2} = I(\lambda,m) \,,
\end{equation}
where 
\begin{equation}
	I(\lambda,m) = \int_{-\pi}^\pi \int_{-\pi}^\pi \frac{\dd q_1 \dd q_2}{(2\pi)^2} \, \frac{1}{\lambda^2 + h^2} \,,
\end{equation}
with
\begin{equation}
	h^2 \coloneqq \frac14 (\sin^2 q_1 + \sin^2 q_2 + (2 + m - \cos q_1 - \cos q_2)^2) \,.
\end{equation}
For given values of $m$ and $W$, the solution of Eq.~\eqref{eq:integral_eq_sp} for $\lambda$  exists, if and only if the graph of $I(\lambda,m)$ intersects the horizontal line $2/W^2$. The integral $I(\lambda,m)$ decreases monotonically with increasing $\lambda$, therefore, a solution exists as long as the global maximum of the integral, $I(0,m)$, is at least as large as $2/W^2$. We therefore encounter the critical disorder strength
\begin{equation} \label{eq:sigma_c_full}
	W_c(m) = \sqrt{\frac{2}{I(0,m)}} \,,
\end{equation}
such that, for $W \geq W_c$, there \textit{exists} a saddle point solution for $\lambda$. Since the integral $I(0,m) \to \infty$ as $m\to0$, the critical disorder strength decays monotonically with the system approaching criticality at $m=0$. Thus, close enough to the critical point, there always exists a solution of Eq.~\eqref{eq:integral_eq_sp} for any given disorder strength $W$. 

\begin{figure}
\centering
\includegraphics[width=0.44\linewidth]{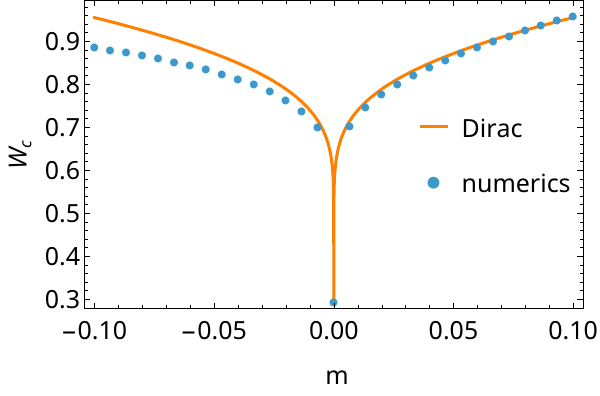} \hspace{1mm}
\includegraphics[width=0.48\linewidth]{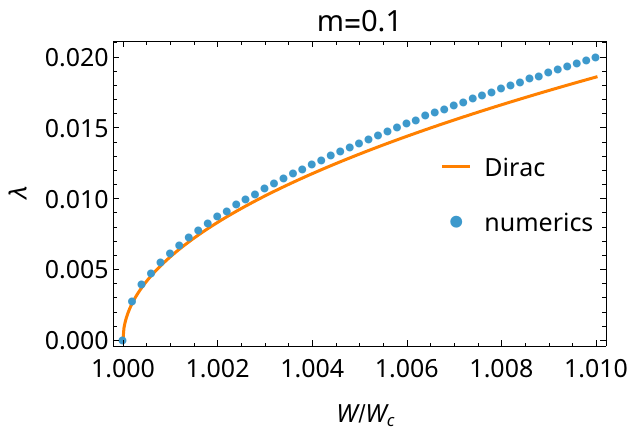}
\caption{Left: the critical disorder strength $W_c (m)$ computed numerically (blue) and analytically in Dirac approximation as in Eq.~\eqref{eq:sigma_c} (orange). 
Right: the solution for $\lambda(W)$ close to the critical disorder strength $W_c$, at $m=0.1$. The blue dots visualize the numerical solution, while the orange line (``Dirac") corresponds to the approximation in Eq.~\eqref{eq:lambda_sqrt}.}
\label{fig:sp_num}
\end{figure}

Since the function $\frac{1}{\lambda^2 + h^2(q)}$ is peaked at the center of the BZ, $q_1 = q_2 = 0$, the integral $I(\lambda,m)$ for $\lambda\ll1$ is well-approximated by the small $q$ linearization (with the Dirac Hamiltonian like in Eq.~\eqref{eq:H_Dirac})
\begin{equation}
\begin{split}
	I_{\tx{lin}}(\lambda,m) 
	&= \frac{1}{(2\pi)^2} \int_{0}^{2\pi} \dd \phi \int_{0}^\pi \dd q  \, \frac{q}{\lambda^2 + \frac14 (m^2 + q^2)} \\
	&\simeq \frac{2}{\pi} \log \frac{\pi}{\sqrt{4\lambda^2 + m^2}} \,,
\end{split}
\end{equation}
where in the last line we approximated $\pi^2 + 4\lambda^2 + m^2 \approx \pi^2$ for $\lambda,m \ll \pi$, and subsequently pulled a power of $2$ out of the logarithm. The global maximum of this function for a given $m$ is then
\begin{equation}
	I_{\tx{lin}}(0,m) \simeq \frac{2}{\pi} \log \frac{\pi}{|m|} \,.
\end{equation}
This leads to an estimate of the critical disorder strength as a function of $m$ as
\begin{equation} \label{eq:sigma_c}
	W_c(m) \simeq \frac{\sqrt \pi}{\sqrt{\log \frac{\pi}{|m|}}} \,.
\end{equation}
As a consistency check, we see that $W_c(m) \to 0$ as $m\to0$ as it should. Note that, in Dirac approximation, $W_c$ is symmetric around $m=0$. This is, however, not the case for the numerical result based on Eq.~\eqref{eq:sigma_c_full}, as can be seen on Fig.~\ref{fig:sp_num}.

Using the approximation $I_{\tx{lin}} (\lambda,m)$ for the value of the integral for weak disorder, i.e., close to the critical disorder strength $W /W_c \sim 1$, we can solve the equation $2/W^2  = I_{\tx{lin}} (\lambda,m)$ for $\lambda$ analytically. The solution reads
\begin{equation} \label{eq:lambda_exponential}
\lambda = \frac12 \sqrt{\pi^2 \exp \rnd{-\frac{2\pi}{W^2}} - m^2} \,, \quad W \geq W_c(m) \,.
\end{equation} 
Expanding in $W$ around the critical disorder strength $W_c$ such that $W /W_c = 1 + \delta w$, we obtain
\begin{equation} \label{eq:lambda_sqrt}
\begin{split}
	\lambda (W) 
	&\simeq |m| \sqrt{\log \frac{\pi}{|m|}} \,\sqrt{\delta w} \\
	&= |m| \sqrt{\log \frac{\pi}{|m|}} \,\sqrt{\frac{W}{W_c} - 1}
\end{split}
\end{equation}

We now analyze the saddle point equation Eq.~\eqref{eq:integral_eq_sp} in the opposite regime: for large values of $\lambda \gg 1$. With the momentum-
dependent term $\frac14 h^2(q_1,q_2) \sim \Ocal(1)$,
the integral $I(\lambda,m)$ is dominated by the $\lambda^{-2}$ contribution for $\lambda \gg 1$. Therefore, in the strong disorder regime, $2/W^2 \ll 1$, we may simply approximate
\begin{equation}
	\lambda(W) \simeq \frac{1}{\sqrt 2} W \,, \quad W \gg 1\,.
\end{equation}
For the intermediate regime of $\lambda \sim \Ocal (1)$, we can use an expansion of the integrand in $I(\lambda,m)$ in orders of $\lambda^{-2}$ up to desired precision.


\section{Moyal expansion of \texorpdfstring{$[\boldsymbol T^{-1},\boldsymbol H] \boldsymbol T$}{[T inverse, H] T}}
\label{app:Moyal}

In the derivation of the continuum action in Sec.~\ref{sec:continuum}, we expand the action in powers of the commutator $[T^{-1},H_0]T$. Dropping the subscript of $H_0$ in the following, we compute the Wigner-Weyl transform 
\begin{equation}
	[T^{-1},H]T \mapsto T^{-1} (x) \star H (q) \star T (x) - H \,,
\end{equation}
with the Moyal $\star$-product defined in Eq.~\eqref{eq:star_def}. Using the fact that $T$-matrices are slowly varying in space, we expand this product in position derivatives to second order, to get
\begin{align}
    T^{-1} (x) \star H (q) \star T (x)
    &= T^{-1} ({x}) \, 
    \e^{\frac{\i}{2}{\larr{\nabla}_x} \cdot {\rarr{\nabla}_q}} \, 
    H ({q}) \, 
    \e^{-\frac{\i}{2} {\larr{\nabla}_q} \cdot { \rarr{\nabla}_x}} \, 
    T ({x})
    \\
    \nonumber
    &\simeq T^{-1} ({x})
    \rnd{1 + \frac{\i}{2} {\larr \del_{x_\mu}}
    {\rarr \del_{q_\mu}} - \frac{1}{8} {\larr \del_{x_\mu}} {\larr \del_{x_\nu}} {\rarr \del_{q_\mu}} {\rarr \del_{q_\nu}} } 
    H ({q}) 
    \\
    \nonumber
    &\hspace{4.4em} \rnd{1 - \frac{\i}{2} {\larr \del_{x_\mu}}
    {\rarr \del_{q_\mu}} - \frac{1}{8} {\larr \del_{q_\mu}} {\larr \del_{q_\nu}} {\rarr \del_{x_\mu}} {\rarr \del_{x_\nu}} } 
    T ({x})
    \\
    \nonumber
    &\simeq H + \frac{\i}{2} {\del_{q_\mu}} H \rnd{ ({\del_{x_\mu}} T^{-1}) T - T^{-1} {\del_{x_\mu}} T } 
    + \frac{1}{4} {\del_{q_\mu} \del_{q_\nu}} H\, {\del_{x_\mu}} T^{-1} {\del_{x_\nu}} T
    \\
    \nonumber
    &\qquad\;
    - \frac{1}{8} {\del_{q_\mu} \del_{q_\nu}} H \rnd{ ({ \del_{x_\mu} \del_{x_\nu}} T^{-1}) T + T^{-1} {\del_{x_\mu} \del_{x_\nu}} T }
    \\
    \nonumber
    &=  H - \i {\del_{q_\mu}} H\,  T^{-1} {\del_{x_\mu}} T
    + \frac{1}{2} {\del_{q_\mu} \del_{q_\nu}} H\, 
    {\del_{x_\mu}} T^{-1} {\del_{x_\nu}} T  \,,
\end{align}
where in the last line we have used that $(\del_{x_\mu} T^{-1}) T = - T^{-1} \del_{x_\mu} T$ due to $T^{-1} T = 1$. 
Defining the variable $\Phi_\mu \coloneqq T^{-1} \del_\mu T$ (Lie-algebra connection) and suppressing the $x$ and $q$ notation inside the derivatives, we thus obtain
\begin{equation}
\begin{split}
	[T^{-1},H]T 
	&\mapsto T^{-1} (x) \star H (q) \star T (x) - H  \\
    &= - \i \del_\mu H\, \Phi_\mu - \frac{1}{2} \del_\mu \del_\nu H\, \Phi_\mu \Phi_\nu \,.
\end{split}
\end{equation}


\section{Derivation of the continuum action}
\label{app:GradientExpansion}

In this appendix, we derive the continuum action in Eq.~\eqref{eq:Scont} from the intermediate result in Eq.~\eqref{eq:SFJ}. Note that this derivation is valid for a large class of general Hamiltonians, since we make few to no assumptions about the explicit form of the Hamiltonian specifying the theory.

\subsection{Gradient and topological terms}

The action to second order in derivatives is given by
\begin{equation}
    S = -\frac{N}{2} (S_1 + S_{12} + S_{22}) \,,
\end{equation}
where
\begin{eqnarray}
        S_1 &\coloneqq &\i \int_{q,x} \tr (G\, \del_\mu H \, \Phi_\mu) \,, \\
        S_{12} &\coloneqq &\frac{1}{2} \int_{q,x} \tr (G\, \del_\mu \del_\nu H\, \Phi_\mu \Phi_\nu) \,, \\
        S_{22} &\coloneqq &\frac{1}{2} \int_{q,x} \tr (G \, \del_\mu H\, \Phi_\mu)^2 \,.
\end{eqnarray}
Here, $G = (\i \lambda \tau_2 -  H)^{-1}$ is the propagator and $\Phi_\mu = T^{-1} \del_\mu T$ a connection constructed from the $T$-matrices and we have again dropped the subscript of $H_0$ for notational convenience.

In the following we will show that the sum of the three terms gives rise to a gradient (or kinetic) term and a topological term 
\begin{equation}
    -(S_1 + S_{12} + S_{22}) = S^\text{grad} + S^\text{top}  \,.
\end{equation}
The first-order term $S_1$ needs special treatment. Naively, being of first order in derivatives, this term seems to vanish. However, this naive supposition is incorrect: the trace over a single Green's function leads to ultraviolet divergent expressions, that is, we are facing a $0 \times \infty$ situation \cite{SpectralFlowAltland2024}. By applying a trick used in, e.g.,
Refs.\ \cite{Moreno2023,SpectralFlowAltland2024}, the number of Green's functions can be doubled to mitigate the UV issues
\begin{equation}
	G = \int_0^\infty \dd \omega \, G_\omega^2 =: \int_\omega G_\omega^2 \,, 
\end{equation}
where $G_\omega = (\i \lambda \tau_2 + \omega - H)^{-1}$ is the frequency dependent propagator. Plugging this identity into $S_1$, we need to again evaluate a Wigner-Moyal product $G \star (\del_\mu H\, \Phi_\mu) \star G$ to second order in derivatives. In doing so, we obtain an additional term of second order in derivatives, which will contribute to the topological action:
\begin{equation}
 	S_1 = \i \int_{q,x} \int_0^{\infty} \dd \omega \, \tr (G_\omega^2 \del_\mu H \,\Phi_\mu)
        + \frac{1}{2} \int_{q,x} \int_0^{\infty} \dd \omega \, \tr ([G_\omega, \del_\nu G_\omega]\, \del_\mu H \,\del_\nu \Phi_\mu) \,.
\end{equation}

From here, it is useful to separate the momentum and position integrals by representing the propagator $G = G^s P^s$ in terms of orthogonal projectors $P^s = \frac{1}{2} (1 + s \tau_2)$, $s=\pm$. We then obtain the expressions for all three terms as
\begin{equation}
    \begin{split}
        S_1 &= \i \int_{q} \sum_s \tr (G^s \del_\mu H) \int_x \tr (P^s \Phi_\mu) \\
        &
        + \frac{1}{2} \int_{q} \int_0^{\infty} \dd \omega \sum_{ss'} \tr (\del_\mu H\, [G_\omega^s, \del_\nu G_\omega^{s'}]) \int_x \tr (P^s P^{s'} \del_\nu \Phi_\mu) \,, \\
        S_{12} &= \frac{1}{2} \int_{q} \sum_s \tr (G^s \del_\mu \del_\nu H) \int_x \tr (P^s \Phi_\mu \Phi_\nu) \,, \\
        S_{22} &= \frac{1}{2} \int_{q} \sum_{ss'} \tr (G^s \del_\mu H G^{s'} \del_\nu H) \int_x \tr (P^s \Phi_\mu P^{s'} \Phi_\nu) \,. \label{eq:G_split}
    \end{split}
\end{equation}

\paragraph{Useful identities.} Due to $\del_\mu (T^{-1} T) = 0$, it holds that $T^{-1} \del_\mu T  = \Phi_\mu = - (\del_\mu T^{-1}) T$. This statement also implies $\del_\mu T^{-1} \del_\nu T = - \Phi_\mu \Phi_\nu$.
Using these transformation rules, one can straightforwardly derive
\begin{equation}
    \del_\mu Q \del_\nu Q = T (\tau_2 \Phi_\mu \tau_2 \Phi_\nu + \Phi_\mu \tau_2 \Phi_\nu \tau_2 - \tau_2 \Phi_\mu \Phi_\nu \tau_2 - \Phi_\mu \Phi_\nu ) T^{-1} \,.
\end{equation}
From this representation, the identities 
\begin{align}
    & \tr (\del_\mu Q \del_\nu Q) = 2 \tr (\tau_2 \Phi_\mu \tau_2 \Phi_\nu) - 2 \tr (\Phi_\mu \Phi_\nu) \,, \label{eq:grad_id1}
    \\
    & \tr (Q \del_\mu Q \del_\nu Q) = - 2 \tr (\tau_2 (\Phi_\mu \Phi_\nu - \Phi_\nu \Phi_\mu))
    \,, \label{eq:grad_id2}
    \\
    & \epsilon_{\mu\nu} \tr (Q \del_\mu Q \del_\nu Q)
    = - 4\epsilon_{\mu\nu} \tr (\tau_2 \Phi_\mu \Phi_\nu)  \label{eq:grad_id3}
     \,
\end{align}
are implied directly.
Armed with these auxiliary identities, we now go through the action $S = S_1 + S_{12} + S_{22}$ term by term and investigate each of them for gradient and topological contributions.

\begin{itemize}
    \item 
    \textbf{$\mathbf{S_1}$-term.}
    For Hamiltonians with a spectrum symmetric under $k_1$- and $k_2$-inversion (such as in our case), the first-order term in $S_1$ vanishes due to the momentum space integration over $\tr (G^s \del_\mu H)$. For the second order term, we use the antisymmetry of $\tr ([G_\omega^s, \del_\nu G_\omega^s] \del_\mu H)$ under $\mu \leftrightarrow \nu$
to decouple the index summation between the momentum and position parts of the term. The reverse use of the Levi-Civita identity $\epsilon_{\mu\nu} \epsilon_{\rho\sigma} = \delta_{\mu \rho} \delta_{\nu \sigma} - \delta_{\mu \sigma} \delta_{\nu \rho}$, followed by the identity $\epsilon_{\mu\nu} \del_\nu \Phi_\mu = \epsilon_{\mu\nu} \Phi_\mu \Phi_\nu$, then produces
	\begin{equation}
	\begin{split}
		S_1 = \frac{1}{4} \int_{q} \int_0^{\infty} \dd \omega \sum_{ss'} \epsilon_{\mu\nu} \tr ([G_\omega^s, 	\del_\nu G_\omega^{s'}] \del_\mu H) \int_x \epsilon_{\rho\sigma} \tr (P^{s} P^{s'} \Phi_\rho \Phi_\sigma) \,. \\
\end{split}
	\end{equation}	
	Due to the Levi-Civita symbol $\epsilon_{\rho\sigma}$ the second trace vanishes for the trivial component of $P^s P^{s'}= \delta_{ss'} P^s = \frac 1 2 \delta_{ss'}(1+s\tau_2)$. For the component proportional to $\tau_2$ we use the identity in Eq.~\eqref{eq:grad_id3} and as a result find a \emph{topological contribution}
    \begin{equation}
        S_1 = -I_1 \int_x \epsilon_{\mu\nu} \tr \rnd{Q \del_\mu Q \del_\nu Q} \,,
    \end{equation}
 	with
 	\begin{equation}
 	\begin{split}
 	I_1 
 	&\coloneqq \frac{1}{32} \int_{q} \int_0^{\infty} \dd \omega \sum_s s \epsilon_{\rho\sigma} \tr \rnd{[G_\omega^s, \del_\sigma G_\omega^s] \del_\rho H} \\
	&= \frac{1}{32} \int_{q} \int_0^{\infty} \dd \omega \sum_s s \epsilon_{\rho\sigma} \tr \rnd{G^s_\omega \del_\rho H  (G^s_\omega)^2 \del_\sigma H - (G^s_\omega)^2 \del_\rho H G^s_\omega \del_\sigma H}
 	\,,
 	\end{split}
 	\end{equation}
 	where in the last line we have used the identity $\del_\sigma G = - G (\del_\sigma G^{-1}) G = G (\del_\sigma H) G$.

    \item 
    \textbf{$\mathbf{S_{12}}$-term.}
Integrating $S_{12}$ in Eq.~\eqref{eq:G_split} by parts in momentum space we obtain
\begin{equation}
     S_{12} = -\frac{1}{2} \int_{q} \sum_s \tr (\del_\mu G^s \del_\nu H) \int_x \tr (P^s \Phi_\mu \Phi_\nu) \,.
\end{equation}
Using $\del_\mu G = - G (\del_\mu G^{-1}) G$, with $\del_\mu G^{-1} = - \del_\mu H$, again, we are left with
\begin{equation} 
    S_{12} = - \frac{1}{2} \int_{q} \sum_s \tr (G^s \del_\mu H G^s \del_\nu H) \int_x \tr (P^s \Phi_\mu \Phi_\nu) \,. \label{eq:anomal}
\end{equation}
This is an anomalous term that will cancel out against a term from the other second order contribution $S_{22}$.

    \item 
    \textbf{$\mathbf{S_{22}}$-term.}
To compute
\begin{equation} \label{eq:S_22_projectors}
    S_{22} = \frac{1}{2} \int_{q} \sum_{ss'} \tr (G^s \del_\mu H G^{s'} \del_\nu H) \int_x \tr (P^s \Phi_\mu P^{s'} \Phi_\nu) \,
\end{equation}
we expand
\begin{align}
    \tr (P^s \Phi_\mu P^{s'} \Phi_\nu) 
    &= \frac{1}{4} \tr \rnd{ss' \tau_2 \Phi_\mu \tau_2 \Phi_\nu + \Phi_\mu \Phi_\nu + s \tau_2 \Phi_\mu \Phi_\nu + s' \tau_2 \Phi_\nu \Phi_\mu} \,
\end{align}
and use (variants of) the auxiliary identities in Eq.~\eqref{eq:grad_id1} and \eqref{eq:grad_id2}, namely,
\begin{equation}
\begin{split}
    &(1) \quad
    \tr(\tau_2 \Phi_\mu \tau_2 \Phi_\nu) = \frac{1}{2} \tr (\del_\mu Q \del_\nu Q) + \tr (\Phi_\mu \Phi_\nu) \,,\\
    &(2) \quad
    \tr (\tau_2 \Phi_\nu \Phi_\mu) = \frac{1}{2} \tr (Q \del_\mu Q \del_\nu Q) + \tr (\tau_2 \Phi_\mu \Phi_\nu) \,,
\end{split}
\end{equation}
to obtain
\begin{align}
    \tr (P^s \Phi_\mu P^{s'} \Phi_\nu) \nonumber
    &= \frac{1}{8} ss' \tr(\del_\mu Q \del_\nu Q) + \frac{1}{8} s' \tr (Q \del_\mu Q \del_\nu Q) \\
    &+ \frac{1}{4} (1 + ss') \tr (\Phi_\mu \Phi_\nu) + \frac{1}{4} (s + s') \tr (\tau_2 \Phi_\mu \Phi_\nu) \,.
\end{align}
We then use the fact that $1 + ss' = 2 \delta_{ss'}$ and $s + s' = 2 s \delta_{ss'}$
to obtain
\begin{equation}
    \tr (P^s \Phi_\mu P^{s'} \Phi_\nu)
    = \frac{1}{8} ss' \tr(\del_\mu Q \del_\nu Q) + \frac{1}{8} s' \tr (Q \del_\mu Q \del_\nu Q)
    + \delta_{ss'} \tr (P^s \Phi_\mu \Phi_\nu)\,.
\end{equation}

Plugging this result into Eq.~\eqref{eq:S_22_projectors}, we arrive at
\begin{align} \label{eq:S_22_final}
    S_{22} &= 
    \frac{1}{16}\int_{q} \sum_{ss'} ss' \tr (G^s \del_\mu H G^{s'} \del_\nu H) \int_x \tr(\del_\mu Q \del_\nu Q)
    \\ &\; \nonumber
    + \frac{1}{16} \int_{q} \sum_{ss'} s' \tr (G^s \del_\mu H G^{s'} \del_\nu H) \int_x \tr (Q \del_\mu Q \del_\nu Q)
    \\ &\;
    +  \underbrace{\frac{1}{2} \int_{q} \sum_{s} \tr (G^s \del_\mu H G^{s} \del_\nu H) \tr (P^s \Phi_\mu \Phi_\nu)}_{-S_{12}} \,. \nonumber
\end{align}


The last term in Eq.~\eqref{eq:S_22_final} cancels the anomalous term $S_{12}$ in Eq.~\eqref{eq:anomal}. The first term in Eq.~\eqref{eq:S_22_final} gives rise to a 
\emph{gradient} term. To see this, we use $ss' = \delta_{ss'} - \delta_{s,-s'}$, as well as $G^s - G^{-s} = 2 \i \Im{G^s}$, and cast it into the form
\begin{equation}
    S_{22}^{\tx{grad}} [Q] = - \sum_{\mu\nu} I_{\mu\nu} \int_x \tr(\del_\mu Q \del_\nu Q) \,,
\end{equation}
with the momentum integrals
\begin{equation}
    I_{\mu\nu} = -\frac{\i}{8} \int_{q} \sum_{s} \tr (G^s \del_\mu H (\Im G^{s}) \del_\nu H).
\end{equation}
The second term in Eq.~\eqref{eq:S_22_final} gives rise to another \emph{topological contribution}. To derive it we use the antisymmetry of $\tr (Q \del_\mu Q \del_\nu Q)$ and observe that contributions with $s=s'$ or $\mu=\nu$ vanish. The remaining ones can be cast into the form
\begin{equation}
    S_{22}^{\tx{top}} [Q] = - I_2 \int_x \epsilon_{\mu\nu} \tr (Q \del_\mu Q \del_\nu Q) \,,
\end{equation}
where the momentum integral is defined as
\begin{equation}
    I_2 = \frac{1}{32} \int_{q} \sum_{s} s \epsilon_{\rho\sigma} \tr (G^s \del_\rho H G^{-s} \del_\sigma H) \,.
\end{equation}
\end{itemize}


In summary, for the total action (cf.\ Eq.~\eqref{eq:Scont})
\begin{equation*}
	S = - \frac N 2 (S_1 + S_{12} + S_{22}) = \frac N 2 (S^{\tx{grad}} + S^{\tx{top}}) \,,
\end{equation*}
we thus have obtained the kinetic and topological contributions 
\begin{equation}
    \begin{split}
        S^{\tx{grad}} [Q] &= \sum_{\mu\nu} I_{\mu\nu} \int_x \tr (\del_\mu Q \del_\nu Q) \,, \\
        S^{\tx{top}} [Q] &= (I_1 + I_2) \int_x \epsilon_{\mu\nu} \tr (Q \del_\mu Q \del_\nu Q) 
    \end{split}
\end{equation}
where the momentum integrals are given by
\begin{align}
    I_{\mu\nu} &\coloneqq -\frac{\i}{8} \int_{q} \sum_{s} \tr (G^s \del_\mu H (\Im G^{s}) \del_\nu H) \,, \\
    I_1 &\coloneqq \frac{1}{32} \int_{q} \int_0^{\infty} \dd \omega \sum_s s \epsilon_{\rho\sigma} \tr \rnd{G^s_\omega \del_\rho H (G^s_\omega)^2 \del_\sigma H - (G^s_\omega)^2 \del_\rho H G^s_\omega \del_\sigma H} \,, \\
    I_2 &\coloneqq \frac{1}{32} \int_{q} \sum_{s} s \epsilon_{\rho\sigma} \tr (G^s \del_\rho H G^{-s} \del_\sigma H) \,.
\end{align}

\subsection{Momentum integrals} \label{app:momentum}

In order to simplify the momentum integrals found as prefactors in the terms above, we represent the Hamiltonian in its spectral decomposition $H = \sum_\alpha \eps_\alpha \ketbra \alpha \alpha$. Plugging in this decomposition allows us to work in terms of the expectation values $G^s_\alpha = \braket{\alpha| G^s | \alpha} = 1/(\i \lambda s -  \eps_\alpha) \in \C$. We may then take the real and imaginary parts of $G^s_\alpha$ explicitly
\begin{align}
    G^s_\alpha &= \frac{1}{\i \lambda s -  \eps_\alpha} = - \frac{\eps_\alpha}{\eps_\alpha^2 + \lambda^2} - \i \frac{s\lambda}{\eps_\alpha^2 + \lambda^2} \\
    &\overset{\lambda \to 0^+}{\simeq} -\frac{1}{\eps_\alpha} - \i s \pi \delta(\eps_\alpha),
    \nonumber
\end{align}
where in the second line we have used the fact that, for weak enough disorder, the imaginary part of $G^s_\alpha$ is peaked sharply enough to be approximated by a Dirac-delta distribution. Apart from this, some useful identities are
\begin{equation} \label{eq:G_alpha_ids}
\begin{split}
    &(1) \quad 
    G^s_\alpha G^s_\beta = \frac{1}{\eps_\alpha - \eps_\beta} (G^s_\alpha - G^s_\beta) \,, \\
    &(2) \quad 
    G^s_\alpha G^{-s}_\beta = \frac{1}{\eps_\alpha - \eps_\beta + 2 \i s \lambda} (G^{s}_\alpha - G^{-s}_\beta)
    \,, \\
    &(3) \quad
    G^s_\beta - G^{-s}_\beta = 2 \i \Im G^s_\beta = - 2 \i s \lambda \, G^s_\beta G^{-s}_\beta
    \,, \\
    &(4) \quad
    \sum_s s \int_0^\infty \dd \omega\, {G^s_\alpha (\omega)}^2 = \sum_s s \, G^s_\alpha (\omega=0) \overset{\lambda \to 0^+}{\simeq} - 2 \pi \i \,\delta(\eps_\alpha) \,, \\
    &(5) \quad
    \sum_s s \int_\omega G^s_\alpha (\omega) \overset{\lambda \to 0^+}{\simeq} -2 \pi \i \int_0^\infty \dd \omega \, \delta(\omega - \eps_\alpha) = - 2 \pi \i \,\Theta(\eps_\alpha) \,, \\
    &(6\txr{a}) \quad
    \braket{\alpha | \del_\mu H | \beta} = \del_\mu \eps_\alpha \delta_{\alpha \beta} + (\eps_\alpha - \eps_\beta) \braket{\del_\mu\alpha| \beta} \,, \\
    &(6\txr{b}) \quad
    \braket{\beta | \del_\mu H | \alpha} = \del_\mu \eps_\alpha \delta_{\alpha \beta} + (\eps_\alpha - \eps_\beta) \braket{\beta | \del_\mu \alpha} \,,
\end{split}
\end{equation}
where in the last two lines we have used $\braket{\del_\mu\alpha | \beta} = - \braket{\alpha | \del_\mu \beta}$. 

\paragraph{Gradient term.} Using the identities in Eq.~\eqref{eq:G_alpha_ids}, it is straightforward to obtain
\begin{equation}
\begin{split}
    I_{\mu\nu} 
    &= -\frac{\i}{8} \int_{q} \sum_{s} \tr (G^s \del_\mu H (\Im G^{s}) \del_\nu H) \\
    &\overset{\lambda \to 0^+}{\simeq} \frac{4 \pi^2}{16} \int_{q} \sum_{\alpha \beta} \delta (\eps_\alpha) \delta (\eps_\beta)  \braket{\alpha | \del_\mu H | \beta} \braket{\beta | \del_\nu H | \alpha} \\
    &\overset{\tx{6a,\,6b}}{=} \frac{1}{16} \int_{\tx{BZ}} \sum_\alpha \dd^2 q \, \delta (\eps_\alpha) \, \del_\mu \eps_\alpha \del_\nu \eps_{\alpha} \,.
\end{split}
\end{equation}
For any Hamiltonian whose eigenvalues possess inversion symmetry in at least one direction $\eps_\alpha (-k_\mu) = \eps_\alpha (k_\mu)$, the integrals $I_{\mu\nu}$ vanish for $\mu \neq \nu$, as long as the BZ is symmetric under $k_\mu \mapsto -k_\mu$.

To make things more concrete, we evaluate the integral explicitly for the 2-band approximation of $H_0$, i.e., the Haldane-Chern insulator $H^{(2)}$ in Eq.~\eqref{eq:Haldane_Ham} \emph{without} relying on the limit $\lambda \to 0$. In this case we have the explicit form of the propagator \begin{equation}
	G^s = D (\i s \lambda + h_a \sigma_a), \quad D = -1 / (\lambda^2 + h^2) \,
\end{equation}
and the integral becomes
\begin{equation}
	I_{\mu\nu} = -\frac{1}{16} \int_{q} \sum_{s} \tr (G^s \del_\mu H (G^{s} - G^{-s}) \del_\nu H) = - \int_{q} D^2 \sum_{s} \tr \Big((\i s \lambda + h_a \sigma_a) \del_\mu H (2 \i s \lambda) \del_\nu H \Big) \,.
\end{equation}
Taking the sum over $s$ and the trace we again find that $I_{\mu\nu}=0$ for $\mu \neq \nu$ and 
\begin{equation}
	I_{\mu\mu} = \frac{\lambda^2}{2} \int_q D^2 \del_\mu h_a \del_\mu h_a \,, 
\end{equation}
where the sum over $a$ is implied, but there is no sum over $\mu$.
Since
\begin{equation}
\begin{split}
	\{ \del_1 h_a \} &= \frac 1 2 (\cos q_1 , \sin q_1, 0 )^\T \,, \quad \tx{and}\\
	\{ \del_2 h_a \} &= \frac 1 2 (0, \sin q_2, \cos q_2 )^\T,
\end{split}
\end{equation}
it holds that $\del_1 h_a \del_1 h_a  = \del_2 h_a \del_2 h_a = \frac14$. We thus find $I_{11}=I_{22}$ and obtain the coupling constant
\begin{equation} \label{eq:g-twoband}
	g \coloneqq I_{11} = I_{22} =  \frac{\lambda^2}{8} \int_{\tx{BZ}} \frac{\dd^2 q}{(2\pi)^2} \,  \frac{1}{(\lambda^2 + h^2)^2} \,.
\end{equation}
Now, upon factoring out $g$, the kinetic/gradient term reads
\begin{equation}
	S^{\tx{grad}} [Q] = g \int_x \tr (\del_\mu Q \del_\mu Q) \,.
\end{equation}
For weak disorder and at low energy, we can view the HC Hamiltonian in Dirac approximation, which yields the approximate value
\begin{equation} \label{eq:g-dirac}
	g \simeq \frac{\lambda^2}{8\pi (\lambda^2 + m^2/4)}< \frac{1}{8 \pi} \ll 1 \,,
\end{equation}
which quantifies the argument for the bare conductance of a one-layer system being small ($\sim\Ocal(1)$), illustrating the necessity of the large control parameter $N$.

\paragraph{Topological term.} For the topological term, we write
\begin{equation} \label{eq:sum_ints_f}
	I_1 + I_2 =  \int_q \sum_{\alpha,\beta} f_{\alpha\beta} \epsilon_{\rho\sigma} \braket{\alpha | \del_\rho H | \beta} \braket{\beta| \del_\sigma H | \alpha}
\end{equation}
with
\begin{equation}
\begin{split}
f_{\alpha\beta} 
&\ev 
\frac{1}{32} \sum_s s \rnd{ \int_\omega \rnd{G^s_\beta(\omega)^2 G^s_\alpha (\omega) - G^s_\alpha(\omega)^2 G^s_\beta (\omega)} + G^{s}_\alpha G^{-s}_\beta } 
\,.
\end{split}
\end{equation}
First, we use identity (1) in Eq.~\eqref{eq:G_alpha_ids} to rewrite
\begin{equation}
	(G^s_\beta)^2 G^s_\alpha - (G^s_\alpha)^2 G^s_\beta= - \frac{1}{\eps_\alpha - \eps_\beta} \rnd{(G^s_\alpha)^2 + (G^s_\beta)^2}
	+ \frac{2}{(\eps_\alpha - \eps_\beta)^2} \rnd{G^s_\alpha - G^s_\beta} \,.
\end{equation}
Then, using identities (4) and (5) in Eq.~\eqref{eq:G_alpha_ids}, we obtain
\begin{equation} \label{eq:I_1_frequency_ints}
\begin{split}
	\sum_s s \int_\omega \rnd{G^s_\beta(\omega)^2 G^s_\alpha (\omega) - G^s_\alpha(\omega)^2 G^s_\beta (\omega)}
	\simeq & \frac{2 \pi \i}{\eps_\alpha - \eps_\beta} ( \delta(\eps_\alpha) + \delta(\eps_\beta) ) \\
	&- \frac{4\pi \i}{(\eps_\alpha - \eps_\beta)^2} (\Theta(\eps_\alpha) - \Theta(\eps_\beta)) 
	\,.
\end{split}
\end{equation}
Going 
further, 
the 
use 
of 
identity 
(2) 
in Eq.~\eqref{eq:G_alpha_ids} delivers
\begin{equation}
	\sum_s s G^s_\alpha G^{-s}_\beta \simeq - \frac{2 \pi \i}{\eps_\alpha - \eps_\beta} (\delta (\eps_\alpha) + \delta(\eps_\beta)) \,,
\end{equation}
which is exactly the negative of the first term in Eq.~\eqref{eq:I_1_frequency_ints}. We, therefore,  obtain
\begin{equation}
	f_{\alpha\beta} \simeq -\frac{\i \pi}{8} \frac{1}{(\eps_\alpha - \eps_\beta)^2} (\Theta(\eps_\alpha) - \Theta(\eps_\beta)) \,.
\end{equation}
Plugging this expression into Eq.~\eqref{eq:sum_ints_f} and using the antisymmetry under the exchange of $\alpha \leftrightarrow \beta$, we obtain
\begin{align}
	I_1 + I_2 
	&\simeq -\frac{\i \pi}{4} \int_q \sum_{\alpha,\beta} \frac{1}{(\eps_\alpha - \eps_\beta)^2} \Theta(\eps_\alpha)
	\,\epsilon_{\rho\sigma} \braket{\alpha | \del_\rho H | \beta} \braket{\beta| \del_\sigma H | \alpha} \\
    \nonumber
	&\overset{\tx{6a,\,6b}}{=}
	-\frac{\pi}{4} \int_q \sum_\alpha \Theta(\eps_\alpha) \,\i \epsilon_{\rho\sigma} \braket{\del_\rho \alpha | \del_\sigma \alpha} \\
	&= -\frac{1}{16 \pi} \sum_\alpha \int_{\tx{BZ}} \dd^2 q \, \Theta(\eps_\alpha) \,\Omega_\alpha (q)
	\,,
     \nonumber
\end{align}
where $\Omega_\alpha (q) = \i \epsilon_{\rho\sigma} \braket{\del_\rho \alpha | \del_\sigma \alpha}$ is the associated Berry curvature and the fraction of the total Berry flux carried by the bands above zero energy defines the \emph{topological angle} 
\begin{equation} \label{eq:theta-berry}
	\vartheta = -\sum_\alpha \int_{\tx{BZ}} \dd^2 q \, \Theta(\eps_\alpha) \,\Omega_\alpha (q) \,.
\end{equation} 
This illustrates how band topology is encoded in the coupling of the topological term of the effective field theory,
\begin{equation}
	S^\text{top}[Q]=\frac{\vartheta}{16 \pi} \int_x \epsilon_{\mu\nu} \tr (Q \del_\mu Q \del_\nu Q)\;.
\end{equation}
Equation~\eqref{eq:theta-berry} has the same structure as the Berry-curvature integral defining the TKNN invariant (named after Thouless, Kohmoto, Nightingale and den Nijs  \cite{TKNN1982}) which is the total Chern number $C$ (summed over the Chern numbers of occupied bands \cite{QiZhang2011review})
\begin{equation}
 	C = \frac{1}{2\pi}\sum_{\alpha\in\mathrm{occ}}\int_{\mathrm{BZ}} \dd^2q \; \Omega_\alpha(q)\in\mathbb{Z}\,.
\end{equation}
In our class~D (BdG) setting the step function $\Theta(\eps_\alpha)$ instead selects positive-energy states. For a fully gapped particle-hole symmetric spectrum this implies $\vartheta\in 2\pi\,\mathbb{Z}$ whenever the Chern number is quantized. In particular, a comparison with Table~1 in Ref.~\cite{wille2023topodual} shows that the Chern numbers of the positive energy bands are related to that of the negative bands by a factor of minus one and thus $\vartheta= 2\pi \, C$.
 Unlike $C$, which is defined for the clean band structure, $\vartheta$ appears as a coupling of the long-distance $\sigma$-model and weights field configurations by their winding number.

To summarize, we state the action of the $N$-level system 
\begin{equation}
S=\frac{N}{2}(\Sgrad+\Stop) 
\end{equation}
which is
\begin{equation}
	S[Q] = \frac{N}{2} \rnd{ g \int_x \tr (\del_\mu Q \del_\mu Q) + \frac{\vartheta}{16 \pi} \int_x \epsilon_{\mu\nu} \tr (Q \del_\mu Q \del_\nu Q) } \,,
\end{equation}
with the coupling constants 
\begin{equation} \label{eq:bare-couplings}
\begin{split}
	g =\frac{ \lambda^2}{8} \int_{\tx{BZ}} \frac{\dd^2 q}{(2\pi)^2} \,\frac{1}{(\lambda^2 + h^2)^2} \,, \\
	\vartheta = -  \sum_\alpha \int_{\tx{BZ}} \dd^2 q \, \Theta(\eps_\alpha) \,\Omega_\alpha (q) \,.
\end{split}
\end{equation}


\end{appendix}






\bibliography{refs}


\end{document}